%% file: paper98e.tex
\documentclass[11pt,a4paper]{article}
\usepackage[british]{babel}
\usepackage{a4wide}
\usepackage[T1]{fontenc}
\usepackage[latin1]{inputenc}
\usepackage[dvips]{graphicx}
\usepackage{psfrag}
\usepackage{charter} % utopia in orige, ma arXiv lo da` per obsoleto  
\usepackage{aa-style}
\input{aa-macros}

 \newcommand{\pinv}{^{(-1)}}
\newcommand{\der}[3][\hphantom{}]{
     \displaystyle\frac{\diff^{#1}{#2}}{\diff{#3}}}
\newcommand{\M}{\hphantom{-}}
\title{\Large\scshape\bfseries
   Statistical applications \linebreak[3]
   of the multivariate skew-normal distribution}
\author{
A.\,Azzalini \\
   Department of Statistical Sciences\\
   University of Padua, Italy\\
   e-mail:~{\tt azzalini@mailhost.stat.unipd.it}
   \and 
A.\,Capitanio\\
   Department of Statistical Sciences\\
   University of Bologna, Italy \\
   e-mail:~{\tt capitani@stat.unibo.it}
   }
\date{February 1998 \\
  (revision of December 1998, with amendment of September 2001)\\
  ~\\
  {\sl This is the full-length version of the paper with the same title \\
  which appears in:} {\em J.\,Roy.\,Statist.\,Soc., series B, vol.61
  (1999), no.\,3}
  }
\begin{document}
\maketitle
\vskip 5ex
\centerline{\bf Summary}
Azzalini \& Dalla Valle (1996) have recently discussed 
the multivariate skew-normal distribution which extends the class of
normal distributions by the addition of a shape parameter.
The first part of the present paper examines further probabilistic 
properties of the distribution, with special emphasis on aspects
of statistical relevance.
Inferential and other statistical issues are discussed in the
following part, with applications to some
multivariate statistics problems, illustrated by numerical examples.
Finally, a further extension  is described which introduces a skewing 
factor of an elliptical density.

%==============================================================================
\clearpage

\section{Introduction}  \label{s:intro}

There is a general tendency in the statistical literature towards 
more flexible methods, to represent features of the data
as adequately as possible and reduce unrealistic assumptions.
For the treatment of continuous multivariate observations within 
a parametric approach, one aspect which has been little affected 
by the above process is the overwhelming role played by the assumption
of normality which underlies most methods for multivariate analysis.
A major reason for this state of affairs is certainly the unrivaled
mathematical tractability of the multivariate normal distribution, in
particular its simplicity when dealing with fundamental operations like
linear combinations, marginalization and conditioning, and indeed its
closure under these operations.

From a practical viewpoint,  the most commonly adopted approach is
transformation of the variables to achieve multivariate normality, 
and in a number of cases this works satisfactorily. 
There are however also problems:
(i) the transformations are  usually on each component separately, 
    and achievement of joint normality is only hoped for;
(ii) the transformed    variables are more difficult to deal with  
   as for interpretation, especially when each variable is
   transformed using a different function;
(iii) when multivariate homoscedasticity is required, this often
   requires a  different transformation from the one for normality.

Alternatively,  there exist several other parametric classes of 
multivariate distributions to choose from, although the choice is
not as wide as in univariate case; 
many of them are reviewed  by Johnson \& Kotz (1972).
A special mention is due to the hyperbolic distribution and its 
generalized version, which form a
very flexible and mathematically fairly tractable parametric class;
see Barndorff-Nielsen \& Bl{\ae}sild (1983) for a summary account,
and Bl{\ae}sild (1981) for a detailed treatment of the 
bivariate case and a numerical example.

As for extensions of  distribution theory of classical statistical methods,
the direction which seems to have been  explored more systematically 
in this context is the extension of distribution theory of traditional
sample statistics to the case of elliptical distribution of the 
underlying population;
elliptical distributions represent a natural extension of the concept of 
symmetry to the multivariate setting. 
The main results in this area are summarized by 
Fang, Kotz \& Ng (1990); see also Muirhead (1982, chapters 1 and 8).

Except for data transformation, however, no alternative method to the
multivariate normal distribution has been adopted for regular use
in applied work, within the framework considered here of a parametric
approach to handle continuous multivariate data.

The present paper examines a different direction of the above
broad problem, namely the possibility to extend some of the classical
methods to the class of multivariate skew-normal distributions which 
has recently been discussed by Azzalini \& Dalla Valle~(1996).
This distribution represents a mathematically tractable extension
of the multivariate normal density with the addition of a parameter to
regulate skewness.

We aim at demostrating that this distribution achieves a reasonable
flexibility in real data fitting, while it maintains a
number of convenient formal properties of the normal one. 
In particular, associated distribution theory of linear and quadratic 
forms remains largely valid.

More specifically, the targets of the paper are as follows:
(a) to extend the analysis of the probabilistic aspects of the
  multivariate skew-normal distribution,
  especially when they reproduce or resemble similar properties of the
  normal distribution;
(b) to examine the potential applications of this distribution in
   statistics, with special emphasis on multivariate analysis.
Correspondingly, after a summary of known results about the distribution,
sections \ref{s:transforms}, \ref{s:cum} and \ref{s:3param} deal with
distribution of linear and quadratic forms of skew-normal variates,
and other probabilistic aspects; 
sections \ref{s:stat} and \ref{s:multivar} deal with issues of more 
direct statistical relevance,
with some numerical examples for illustration.
In addition,
section~\ref{s:elliptical} sketches an additional level of generalization
by introducing a skew variant of elliptical densities.

%============================================================================

\section{The multivariate skew-normal distribution}   \label{s:msn}

We first recall the definition and a few key properties of the
distribution, as given by Azzalini \& Dalla Valle~(1996) except for
re-arrangement of the results.
A $k$-dimensional random variable $Z$ is said to have a multivariate
skew-normal distribution if it is continuous with density function
\begin{equation}
         2 \phi_k(z;\Omega)\, \Phi(\alpha\T z) ,  \qquad (z\in\Real^k),
                                                        \label{f:dens}
\end{equation}
where $\phi_k(z;\Omega)$ is the $k$-dimensional normal density with zero
mean and correlation matrix $\Omega$,  $\Phi(\cdot)$ is the $N(0,1)$
distribution function, and $\alpha$ is a $k$-dimensional vector.
For simplicity, $\Omega$ is assumed to be of full rank.

When $\alpha=0$, (\ref{f:dens}) reduces to the $N_k(0,\Omega)$ density.
We then refer to $\alpha$ as a `shape parameter', 
in a broad sense, although the actual shape is regulated in a more
complex way, as it will emerge in the course of the paper.

The above density does not allow location and scale parameters.
Clearly, these are essential in practical statistical work, but we
defer their introduction until later, to keep notation simple as 
long as possible.

The matrix $\Omega$ and the vector $\alpha$ appearing in (\ref{f:dens})  
were defined  in Azzalini \& Dalla Valle (1996) as functions
of other quantities, namely another correlation matrix $\Psi$ and a vector 
$\lambda\in\Real^k$; hence a member of the parametric family was
identified by the pair $(\lambda,\Psi)$. 
It is in fact possible to identify the member of the family 
directly by the pair $(\alpha, \Omega)$; i.e.\ this pair provides
an equivalent parametrization of the class of densities.
The proof of this fact is of purely algebraic nature, and it is given
in an appendix, together with some related results.
For the purposes of the present paper, this parametrization appears
preferable and we shall adopt  the notation  
\[   Z\sim \SN_k(\Omega,\alpha)   \]
to indicate that $Z$ has density function (\ref{f:dens}).

% By simple algebraic transformation of (2.13) of Azzalini \& Dalla
% Valle~(1996), 
The cumulant generating function is
\begin{equation}
    K(t) = \log M(t) = \half t\T \Omega t + \log\{2\,\Phi(\delta\T t)\} 
                                                \label{f:cum}
\end{equation}
where
\begin{equation}
   \delta = \frac{1}{\radice{1+\alpha\T \Omega\alpha}} \Omega\alpha.
                                                \label{f:delta}
\end{equation}
Hence the mean vector and the variance matrix are 
\begin{equation}
   \mu_z   = \E{Z} = \radice{2/\pi}\,\delta, \qquad
   \var{Z} = \Omega-\mu_z\,\mu_z\T.              \label{f:E+var}
\end{equation}

The following result provides a  stochastic representation of $Z$,
useful for computer generation of random numbers and for theoretical
purposes.
\begin{proposition}  \label{th:stoch-repr}
Suppose that 
\[   \pmatrix{X_0 \cr X} \sim  N_{k+1}(0,\Omega^*), \quad  
     \Omega^* =  \pmatrix{1    & \delta\T \cr
                       \delta & \Omega} 
\]
where $X_0$ is a scalar component and $\Omega^*$ is a correlation
matrix.  Then
\[  Z=\cases{ X & if $X_0>0$ \cr
             -X & otherwise  }
\]
is $\SN_k(\Omega, \alpha)$ where 
\begin{equation}
   \alpha = 
       \frac{1}{\radice{1-\delta\T\Omega\inv\delta}} \Omega\inv\delta.
    \label{f:alpha}
\end{equation}
\end{proposition}

Also, we shall make repeated use of the Sherman--Morrison--Woodbury 
formula for matrix inversion, which states
\begin{equation}
     (A+UBV)^{-1}=A^{-1}-A^{-1}UB(B+BVA^{-1}UB)^{-1}BVA^{-1}
                                                  \label{f:smw}
\end{equation}
for any conformable matrices, provided the inverses involved exist;
see for instance Rao (1973, exercise~2.9, p.\,33).

%==============================================================================

\section{Linear and quadratic forms} \label{s:transforms}

A key feature of the multivariate normal distribution is its
simplicity to handle linear and quadratics forms.  We now explore the
behaviour of the skew-normal distribution in these cases.

%--------------------------------------------
\subsection{Marginal distributions}

It is implicit in the genesis of the multivariate skew-normal variate, 
as described by Azzalini \& Dalla Valle~(1996), that the marginal 
distribution of a subset of the components of $Z$ is still a
skew-normal  variate. 
In the marginalization operation, the $(\lambda,\Psi)$
parametrization works in a very simple manner, since one only needs to
extract the relevant components of $\lambda$ and $\Psi$. 
With the $(\Omega,\alpha)$ parametrization, specific formulae must
be developed.

\begin{proposition}   \label{th:marginal}
Suppose that $Z\sim \SN_k(\Omega,\alpha)$ and $Z$ is partitioned as
$Z\T=(Z\T_1, Z\T_2)$ of dimensions $h$ and $k-h$, respectively;
denote by
\[  \Omega= \pmatrix{\Omega_{11} & \Omega_{12} \cr
                     \Omega_{21} & \Omega_{22}}, \qquad
    \alpha = \pmatrix{\alpha_1 \cr \alpha_2}
\]
the corresponding partitions of $\Omega$ and $\alpha$.
Then the marginal distribution of $Z_1$ is 
$\SN_h(\Omega_{11},\bar{\alpha}_1)$, where
\[  
   \bar{\alpha}_1 = \frac{\alpha_1+\Omega_{11}\inv\Omega_{12}\alpha_2}%
                    {(1+\alpha\T_2 \Omega_{22\cdot1} \alpha_2)^{1/2}}, \qquad
   \Omega_{22\cdot1} = \Omega_{22}-\Omega_{21} \Omega_{11}\inv \Omega_{12} . 
\]
\end{proposition}
The proof follows from straightforward integration, with the aid of 
Proposition~4 of Azzalini \& Dalla Valle~(1996).

%-------------------------------------------
\subsection{Linear transforms}

\begin{proposition} \label{th:lin1}
If $Z\sim \SN_k(\Omega,\alpha)$, and $A$ is a non-singular $k\times k$ matrix
such that $A\T\Omega A$ is a correlation matrix, then
\[  A\T Z \sim \SN_k(A\T \Omega A, A\inv \alpha). \]
\end{proposition}
The proof follows from standard rule of transformation of random
variables.
The above condition that $A\T \Omega A$ is a correlation matrix is
there for the sake of simplicity of exposition, and it can be removed;
see section \ref{s:3param}.

\begin{proposition}
For a variable $Z\sim \SN_k(\Omega,\alpha)$, there exists a  linear 
transform $Z^*=A^* Z$ such that  $Z^*\sim \SN_k(I_k,\alpha^*)$ where
at most one component of $\alpha^*$ is not zero. 
\end{proposition}
\emph{Proof.~}  
By using the  factorization $\Omega=C\T C$, we first transform $Z$ into 
a variable $Y=(C\T)\inv Z$ such that $Y\sim \SN_k(I_k,C\alpha)$.
Now consider an orthogonal matrix $P$ with one column on the same
direction of $C\alpha$, and define $Z^*=P\T Y$ which fulfills the
conditions. 
\vskip 2ex
The above result essentially defines a sort of `canonical form' whose
components are mutually independent, with a single component `absorbing' 
all asymmetry of the multivariate distribution. 
This linear transformation plays a role similar to  the one which 
converts a multivariate normal variable into a spherical form.
Further, notice that the component transformations of $A^*$ are invertible; 
hence it is possible to span the whole class
$\SN_k(\Omega,\alpha)$ starting from $Z^*$ and applying suitable linear
transformations.
The density of $Z^*$ is of the form 
\[     2\,\prod_{i=1}^k \phi(u_i)\, \Phi(\alpha^*_m u_m)  \]
where 
\begin{equation}
          \alpha_m^*=\radice{\alpha\T\Omega\alpha}   \label{f:alpha*}
\end{equation}
is the only non-zero component of $\alpha^*$.

For the rest of this section, we examine conditions for independence 
among blocks of components of a linear transform $Y=A\T Z$. 
Before stating the main conclusion, we need the following intermediate result.

\begin{proposition}  \label{lin-dist}
Let $Z\sim \SN_k(\Omega,\alpha)$ and $A$ is as in Proposition~\ref{th:lin1}, 
and consider the linear transform
\begin{equation} 
    Y = A\T Z = \pmatrix{Y_1\cr \vdots \cr Y_h} 
      = \pmatrix{A_1\T\cr \vdots\cr A_h\T} Z          \label{f:AZ}
\end{equation}
where the matrices $A_1, \ldots, A_h$ have $m_1,\ldots,m_h$ columns,
respectively. Then
\[ Y_i \sim \SN_{m_i}(\Omega_{Y_i}, \alpha_{Y_i}) \]
where
\[
  \Omega_{Y_i} = A_i\T \Omega A_i,\qquad 
  \alpha_{Y_i} =  \frac{(A_i\T \Omega A_i)\inv A_i\T \Omega \alpha}
                  {\radice{1+\alpha\T(\Omega-\Omega A_i(A_i\T \Omega A_i)\inv 
                       A_i\T \Omega) \alpha}}
\]
\end{proposition} 
\emph{Proof.~}
Without a loss of generality, we consider the case $h=2$ and $i=1$.
Write $A=(A_1, A_2)$ and denote its inverse by 
\[  A\inv= \pmatrix{A_1\pinv \cr
                   A_2\pinv }
\]
where the number of columns of the blocks of $A$ matches the number of rows
of the blocks of $A\inv$.
Since $A A\inv=I_k$, then the identity $A_1 A_1\pinv + A_2 A_2\pinv=I_k$ holds.
On partitioning $A\T \Omega A$ in an obvious way, and  
\[  A\inv \alpha=\pmatrix{A_1\pinv \alpha\cr A_2\pinv\alpha},  \]
the result follows after some algebra by applying Proposition~\ref{th:marginal} 
to the parameters of $A\T Z$, taking into account the above identity.
\par\vskip 2ex

We now turn to examine the issue of independence among blocks of a linear 
transform $A\T Z$ where $A$ satisfies the condition of 
Proposition~\ref{th:lin1}.
To establish independence among the $Y_i$'s, a key role is played by 
the $\Phi(\cdot)$  component in (\ref{f:dens}).
Since $\Phi(u+v)$ cannot be factorized as the product $\Phi(u)\,\Phi(v)$, 
it follows that at most one of the $Y_i$ can be a `proper' skew-normal 
variate, while the others must have the skewness parameter equal to 0, 
hence be regular normal variates, if mutual independence holds.

\begin{proposition}  \label{th:lin2}
If $Z\sim \SN_k(\Omega,\alpha)$, and $A\T\Omega A$ is a positive definite
correlation matrix,
then the variables $(Y_1,\ldots,Y_h)$ defined by (\ref{f:AZ}) are independent
if and only if the following conditions hold simultaneously:
\begin{itemize}
\item[(a)] $A_i\T \Omega A_j=0$ for $i\ne j$,
\item[(b)] $A_i\T \Omega \alpha \ne 0$ for at most one $i$.
\end{itemize}
\end{proposition}
\emph{Proof.~} Prove sufficiency first. By Proposition \ref{th:lin1}
and condition (a),
the joint distribution of $Y$ is $\SN_k(\Omega_Y, \alpha_Y)$ where
\begin{eqnarray*}
\Omega_Y &=& \diag(A_1\T \Omega A_1, \ldots, A_h\T \Omega A_h), \\
\alpha_Y &=& (A\T \Omega A)\inv A\T \Omega \alpha 
         = \pmatrix{(A_1\T \Omega A_1)\inv A_1\T \Omega \alpha \cr 
                      \vdots \cr
                     (A_h\T \Omega A_h)\inv A_h\T \Omega \alpha} \,.
\end{eqnarray*}
If condition (b) is satisfied too,
only one of the blocks of $\alpha_Y$ is not zero. Hence the joint
density can be factorized in obvious manner.\par
To prove necessity, note that if independence holds the density of $Y$
can be factorized as the product of the densities of the $Y_i$'s,
given by Proposition~\ref{lin-dist}. Since the
function $\Phi$ cannot be factorized, only one block of $\alpha_Y$
can be not zero, and $\Omega_Y$ must be a block-diagonal matrix. 
These requirements can be met only if conditions (a) and (b) are satisfied. 

\vskip 2ex
Notice that the parameters of the $Y_i$'s are equal to the corresponding
blocks of $(\Omega_Y,\alpha_Y)$ only if independence holds.

%--------------------------------------------------

\subsection{Quadratic forms}

One appealing feature of the one-dimensional skew-normal distribution
is that the square of a random variate of this kind is a $\chi_1^2$. 
This property carries on in the multivariate case since 
$Z\T \Omega\inv Z \sim \chi_k^2$, irrespectively of $\alpha$. 
These facts are special cases of the more general results presented
below.

\begin{proposition}     \label{quad-dist}
If $Z\sim \SN_k(\Omega,\alpha)$, and $B$ is a symmetric positive semi-definite
$k \times k$ matrix of
rank p such that $B \Omega B = B$, then $Z\T B Z \sim \chi^2_p$.
\end{proposition}
\emph{Proof.} Consider first the case of a random variable 
$Y \sim \SN_p(I_p,\alpha)$. 
Since $Y\T Y = Y\T A A\T Y$ for any orthogonal matrix $A$, 
hence in particular it holds  for 
a matrix having a column on the same direction of $\alpha$, i.e. we are 
considering the canonical form associated to $Y$.
It then follows that $Y\T Y \sim \chi_p^2$ independently of $\alpha$.
\par
In the general case, let us write $B = M M\T$ where $M$ is a full-rank 
$k \times p$ matrix ($p \le k$), and notice that $M\T \Omega M = I_p$ is 
equivalent to $B\Omega B = B$; to see this, it is sufficient to left-multiply
each side of the latter equality by$(M\T M)\inv M\T$ and right-multiply by 
its transpose. Then $Z\T B Z = Y\T Y$ where $Y = M\T Z \sim 
SN_p(I_p, \alpha_Y)$ for some suitable vector $\alpha_Y$.
Therefore the statement holds because $Y\T Y \sim \chi_p^2$.
\begin{corollary}   \label{quad-chi}
If $Z\sim \SN_k(\Omega,\alpha)$, and $C$ is a full-rank $k \times p$ matrix
($p \le k$), then 
\[
     Z\T C(C\T \Omega C)\inv C\T Z \sim \chi_p^2.
\]
\end{corollary}
\begin{proposition}   \label{quad-indip}
If $Z\sim \SN_k(\Omega,\alpha)$, and $B_i$ is a  symmetric positive 
semi-definite $k \times k$ matrix of rank  $p_i \; (i=1, 2, \ldots, h)$ 
such that 
\begin{itemize}
\item[(a)] $B_i \Omega B_j = 0$ for $i \neq j$,
\item[(b)] $\alpha\T \Omega B_i \Omega \alpha \ne 0$ for at most one $i$,
\end{itemize}
then the quadratic forms $Z\T B_i Z\; (i=1, 2, \ldots, h)$ are mutually 
independent.
\end{proposition}
\emph{Proof.~} Similarly to the proof of Proposition \ref{quad-dist},
write $B_i = M_i M_i\T$ where $M_i$ has rank $p_i$. Clearly
the quadratic forms $Z\T B_i Z$ are mutually independent if this
is true for the linear forms $M_i\T Z$. It is easy to see that
$M_i\T \Omega M_j = 0$ is equivalent to $B_i\T \Omega B_j = 0$ for $i \ne j$;
similarly $M_i\T \Omega \alpha \ne 0$ is equivalent to $\alpha\T \Omega 
B_i \Omega \alpha \ne 0$. This completes the proof. 
 
\begin{proposition}[Fisher--Cochran]
If $Z\sim \SN_k(I_k,\alpha)$ and $B_1,\ldots,B_h$ are symmetric
$k\times k$ matrices of rank $p_1,\ldots,p_h$, respectively, 
such that $\sum B_i=I_k$ and $B_i\alpha\ne 0$ 
for at most one choice of $i$,  then the quadratic forms $Z\T B_i Z$
are independent $\chi^2_{p_i}$ if and only if $\sum p_i=k$.
\end{proposition}
\emph{Proof.}
The proof follows the steps of the usual one of Fisher--Cochran
theorem, as given for instance by Rao (1973, p.\,185 ff.),
taking into account Proposition~\ref{quad-indip}
for independence of the quadratic forms, and 
Proposition~\ref{quad-dist} as for their marginal distributions.
\vskip 2ex
It would be possible to develop this section via a different approach, on the 
basis of Proposition~\ref{th:stoch-repr}.  For most of the results, this route 
would offer a simple treatment, but for some others it would be quite
cumbersome, especially for the results about independence of components.
%==============================================================================

\section{Cumulants and indices}  \label{s:cum}

To study higher order cumulants besides those given in Section~\ref{s:msn},
we need some preliminary results about the cumulants of the
half-normal distribution, i.e.\ the distribution of $V=|U|$,
where $U\sim N(0,1)$.  Its cumulant generating function is
\[  K^V(t) =  \half t^2 + \zeta_0(t)   \]
where
\[  \zeta_0(x) = \log(2\Phi(x)).  \]
For later use, define
\[  
   \zeta_m(x) = \frac{\diff^m}{\diff x^m} \zeta_0(x) \qquad (m=1,2,\ldots).
\]
Clearly,  $\zeta_1(x) = \phi(x)/\Phi(x)$; the subsequent derivatives
can  be expressed as functions of the lower order  derivatives, e.g.
\begin{eqnarray*} 
    \zeta_2(x) &=& -\zeta_1(x) \{x + \zeta_1(x)\}, \\
    \zeta_3(x) &=& -\zeta_2(x)\{x+\zeta_1(x)\}-\zeta_1(x)\{1+\zeta_2(x)\}, \\
    \zeta_4(x) &=& -\zeta_3(x)\{x+2\zeta_1(x)\}-2\zeta_2(x)\{1+\zeta_2(x)\},
\end{eqnarray*} 
hence as functions of $\zeta_1(x)$. Computation of $\zeta_m$
at $x=0$ gives the corresponding cumulant $\kappa_m^V$. 
Unfortunately, it is not clear how to obtain a closed or recursive 
formula for the  $\zeta_m(x)$'s.

An alternative route for computing $\kappa^V_m$ is as follows: 
since $V\sim (\chi_1^2)^{1/2}$ then
\[ \E{V^m} = \frac{2^{m/2}}{\sqrt{\pi}} \Gamma\left(\frac{m+1}{2}\right) \]
which admits the recurrence formula
\[      \E{V^m} = (m-1)\E{V^{m-2}},  \qquad (m\geq2).\]
Hence the cumulant $\kappa^V_m$ can be obtained from  the set $\E{V^r}, \,
r=1,\ldots,m$, using well-known results;
see e.g.\ Table 2.1.2 of David, Kendall \& Burton (1966) for expressions
connecting cumulants to moments up to order 8.
In particular, we obtain for $V$ that
\[ 
    \kappa^V_3 = \radice{2/\pi}(4/\pi-1), \quad 
    \kappa^V_4 = 4 (2-6/\pi)/\pi .
\]
% these expressions are in agrement with those of
% Azzalini~(1985) (which reports standardized cumulants).

Returning to cumulant generating function (\ref{f:cum}), its  first 
two derivatives are
\[ 
  \der{K(t)}{t}                 = \Omega t + \zeta_1(x)\delta, \quad 
  \der[2]{K(t)}{t\, \diff t\T}  = \Omega + \zeta_2(x) \,\delta \delta\T
\]
where $x=\delta\T t$, and its evaluation at $t=0$ confirms (\ref{f:E+var}).
Higher order cumulants are obtained from
\[  \frac{\diff^m K(t)}{\diff t_i\, \diff t_j\, \cdots \diff t_r}
        = \zeta_m(x)\, \delta_i \delta_j \cdots \delta_r 
\]
which needs to be evaluated only at $x=0$ where
\[        \zeta_m(x)\big|_{x=0} = \kappa^V_m\]
which can has been obtained as described above.

One use of these expressions is to obtain summary indicators for the
$\SN_k$ distribution. The most popular ones are those introduced by 
Mardia (1970, 1974) to measure multivariate skewness and kurtosis.
In our case, the index of skewness takes the form
\begin{eqnarray*} 
\gamma_{1,k} &=& \beta_{1,k}  
  = (\kappa^V_3)^2 \sum_{rst} \sum_{r's't'} 
       \delta_r \delta_s \delta_t \delta_{r'} \delta_{s'} \delta_{t'}
       \sigma^{rr'}\sigma^{ss'}\sigma^{tt'}\\
  &=&  \left(\frac{4-\pi}{2}\right)^2 \left(\mu_z\T\Sigma\inv\mu_z\right)^3
\end{eqnarray*} 
where $\Sigma=\Omega-\mu_z\mu_z\T=(\sigma_{rs})$ with inverse 
$\Sigma\inv=(\sigma^{rs})$.
Similarly, the index of kurtosis is
\begin{eqnarray*} 
  \gamma_{2,k} &=& \beta_{2,k}-k(k+2) 
       =  \kappa^V_4 \sum_{rstu} \delta_r \delta_s \delta_t \delta_u
                                     \sigma^{rs}\sigma^{tu}\\
       &=&   2(\pi-3) \left(\mu_z\T\Sigma\inv\mu_z\right)^2.
\end{eqnarray*} 
% These expressions have clear connections with the ones
% presented in Azzalini~(1985) for the univariate case.

There exists an alternative multivariate index of skewness discussed in the
literature; see e.g.\  McCullagh (1987, p.40). However this differs from 
$\gamma_{1,k}$ only by a different way of matching the indices of the 
cumulants, but this has no effect in the present case
because of the special pattern of the cumulants of order higher than 2. 
Hence, in our case the two indices of skewness coincide.

Using  (\ref{f:smw}),  one can re-write
\[  \mu_z\T\Sigma\inv\mu_z = 
   \frac{\mu_z\T\Omega\inv\mu_z}{1-\mu_z\T\Omega\inv\mu_z} \]
which allows easier examination of the range of $\mu_z\T\Sigma\inv\mu_z$, by
considering the range of $\delta\T\Omega\inv\delta$.
On using (\ref{f:delta}), we write
\[ 
    \delta\T\Omega\inv\delta 
         = \frac{\alpha\T\Omega\alpha}{1+\alpha\T\Omega\alpha}
         = \frac{a}{1+a}
\]
where $a$ is the square of $\alpha_m^*$, defined by (\ref{f:alpha*}).
Since $a$ spans $[0,\infty)$, then 
\[ \mu_z\T\Sigma\inv\mu_z = \frac{2a}{\pi+(\pi-2)a} \in [0,2/(\pi-2))  \]
and the approximate maximal values for $\gamma_{1,k}$ and $\gamma_{2,k}$ are
0.9905, and 0.869, respectively, in agreement with the univariate case.
Since both  $\gamma_{1,k}$ and $\gamma_{2,k}$ depend of $(\Omega, \alpha)$ 
only via $\alpha_m^*$, this reinforces the role of the latter as the
summary quantity of the distribution shape.

%==============================================================================

\section{Some extensions}                      \label{s:3param}

\subsection{Location and scale parameters}
For the subsequent development of the paper, we need to introduce
location and scale parameters, which have been omitted in the
expression (\ref{f:dens}) of the density of $Z$.
Write then
\begin{equation}
                  Y = \xi + \omega Z           \label{f:Y}
\end{equation} 
where
\[   
   \xi=(\xi_1,\ldots,\xi_k)\T,  \qquad
   \omega=\diag(\omega_1,\ldots,\omega_k)
\]
are location and scale parameters, respectively; the components of
$\omega$ are assumed to be positive.  
The density function of $Y$ is 
\begin{equation}
     2\,\phi_k(y-\xi;\Omega)\,\Phi\{\alpha\T\omega\inv(y-\xi)\}  
                                                      \label{f:densY}
\end{equation}
where 
\[        \Omega= \omega\Omega_z\omega   \]
is a covariance matrix and, from now on, $\Omega_z$ replaces the
symbol $\Omega$ used in the previous sections.
Hence, for instance, (\ref{f:delta}) must now be read with $\Omega$
replaced by $\Omega_z$.  We shall use the notation
\[   Y \sim \SN_k(\xi, \Omega, \alpha )  \]
to indicate that $Y$ has density function (\ref{f:densY}).
In the sequel, we shall also use the notation $\sqrt{A}$ to denote 
the diagonal matrix of the square root of the diagonal elements of 
a positive definite matrix $A$;
hence, for instance, $\omega=\sqrt{\Omega}$.

Earlier results on linear and quadratic forms for $Z$ carry on for $Y$,
apart for some slight complication in the notation.
For instance, for a linear transform $A\T Y$ where $A$ is
a $k\times h$ matrix, a simple extension of Proposition~\ref{lin-dist}
gives
\begin{equation}
          X = A\T Y \sim \SN_h(\xi_X, \Omega_X, \alpha_X )    
                                                 \label{f:lin-transf}
\end{equation}
where
\[  
   \xi_X= A\T\xi, \qquad \Omega_X = A\T \Omega A, \qquad
   \alpha_x = \frac{\omega_X \Omega_X\inv B\T \alpha}
           {\radice{1+\alpha\T(\Omega_z- B \Omega_X\inv B\T)\alpha}}
\]
and
\[  \omega_X=\sqrt{\Omega_X}, \qquad  B = \omega\inv\Omega A.  \]
Similar extensions could be given for other results of 
Section~\ref{s:transforms}.  For later reference, we write the
new form of the cumulant generating function
\begin{equation}
      K(t)=  t\T \xi + \half t\T \Omega t +\log\{2\,\Phi(\delta\T\omega t)\}.
                                       \label{f:K(t)}
\end{equation}

\subsection{Conditional distributions}
Suppose that $Y$ has density function (\ref{f:densY}), and it is 
partitioned in two components, $Y_1$ and $Y_2$, of dimensions $h$ and
$k-h$, respectively, with a corresponding partition for $\xi$,
$\Omega$ and $\alpha$.
To examine the distribution of $Y_2$ conditionally on $Y_1=y_1$, 
write
\[  
 \xi_2^c = \xi_2 + \Omega_{21} \Omega_{11}\inv(y_1-\xi_1), \quad
 \Omega_{22\cdot1} = \Omega_{22}-\Omega_{21}\Omega_{11}\inv\Omega_{12},
   \quad   
 \bar{\alpha}_1 = 
   \frac{\alpha_1+\omega_1 \Omega_{11}\inv\Omega_{12}\omega_2\inv\alpha_2}%
        {(1+\alpha\T_2 \bar{\Omega}_{22\cdot1} \alpha_2)^{1/2}},
\]
where
\[    \omega_1 = \sqrt{\Omega_{11}},   \qquad
      \omega_2 = \sqrt{\Omega_{22}},   \qquad
      \bar{\Omega}_{22\cdot1}= \omega_2\inv\Omega_{22\cdot1}\omega_2\inv.
\]
Here $\xi_2^c$ and $\Omega_{22\cdot1}$ are given by the usual formulae
for the conditional mean and variance of a normal variable, 
and $\bar{\alpha}_1$ is the shape parameter of the marginal distribution 
of $Y_1$. 
After some straightforward computation, it follows that the
cumulant generating function of the conditional distribution is
\[ 
  K_c(t) =  t\T \xi_2^c + \half t\T \Omega_{22\cdot1}t
           + \log\Phi(x_0+{\tilde\delta}_2\T\omega_2 t) -\log\Phi(x_0)
 \]
where
\[ x_0= \bar{\alpha}_1\T \omega_1\inv(y_1-\xi_1) \]
and ${\tilde\delta}_2$ is computed similarly to (\ref{f:delta}),
with $\Omega$ and $\alpha$  replaced by
$\bar{\Omega}_{22\cdot1}$ and $\alpha_2$, respectively.
This gives immediately
\begin{equation}
    \E{Y_2|y_1}   = \xi_2^c + \zeta_1(x_0) \tau,   \qquad
    \var{Y_2|y_1} = \Omega_{22\cdot1} +\zeta_2(x_0) \tau \tau\T
    \label{f:Evar-cond}
\end{equation}
where  $\tau=\omega_2 {\tilde\delta}_2$; 
higher order cumulants of order $m$ are of the form
\[  
    \zeta_m(x_0)\,\underbrace{\tau_r \tau_s \cdots\tau_u}_{
                              m\hbox{\scriptsize~terms}},  \qquad (m>2),
\]
where $\tau_r$ denotes the $r$-th component of $\tau$.

Clearly, $K_c(t)$  is of form (\ref{f:K(t)}).
This special case occurs only if $x_0=0$; this condition is essentially equivalent 
to $\bar{\alpha}_1=0$, i.e.\ $Y_1$ is marginally normal.

The expression of the conditional density in the general case is easily
written down, namely
\begin{equation}   \label{f:sncond}
       \phi_{k-h}(y_2 -\xi_2^c;\Omega_{22\cdot1}) \,      
           \Phi\{\alpha_2\T \omega_2\inv  (y_2 - \xi_2^c)+ x'_0\}/{\Phi(x_0)}
\end{equation}
where $x'_0=\radice{1+\alpha_2\T \,\bar{\Omega}_{22.1}\,\alpha_2}x_0$.
In the case $k-h=1$, this distribution has been  discussed by several
people, including  Chou \& Owen (1984), Azzalini (1985), Cartinhour (1990)
and Arnold {\em et al.} (1993).
From (\ref{f:sncond}), it is easy to see that conditions for independence 
among components are the same of the unconditional case, with 
$\Omega_{22\cdot1}$ and $\alpha_2$ replacing $\Omega$ and $\alpha$, 
confirming again the usefulness of the adopted parametrization.

The shape of (\ref{f:sncond}) depends on a number of ingredients;
however, for most cases, the plot of this density function displays 
a remarkable similarity with the one of the skew-normal density.
This similarity suggests the approximation of the conditional density by
a skew-normal density which matches cumulants up to the third order.

\begin{figure}
\centerline{
  \includegraphics[width=0.49\hsize,height=9cm]{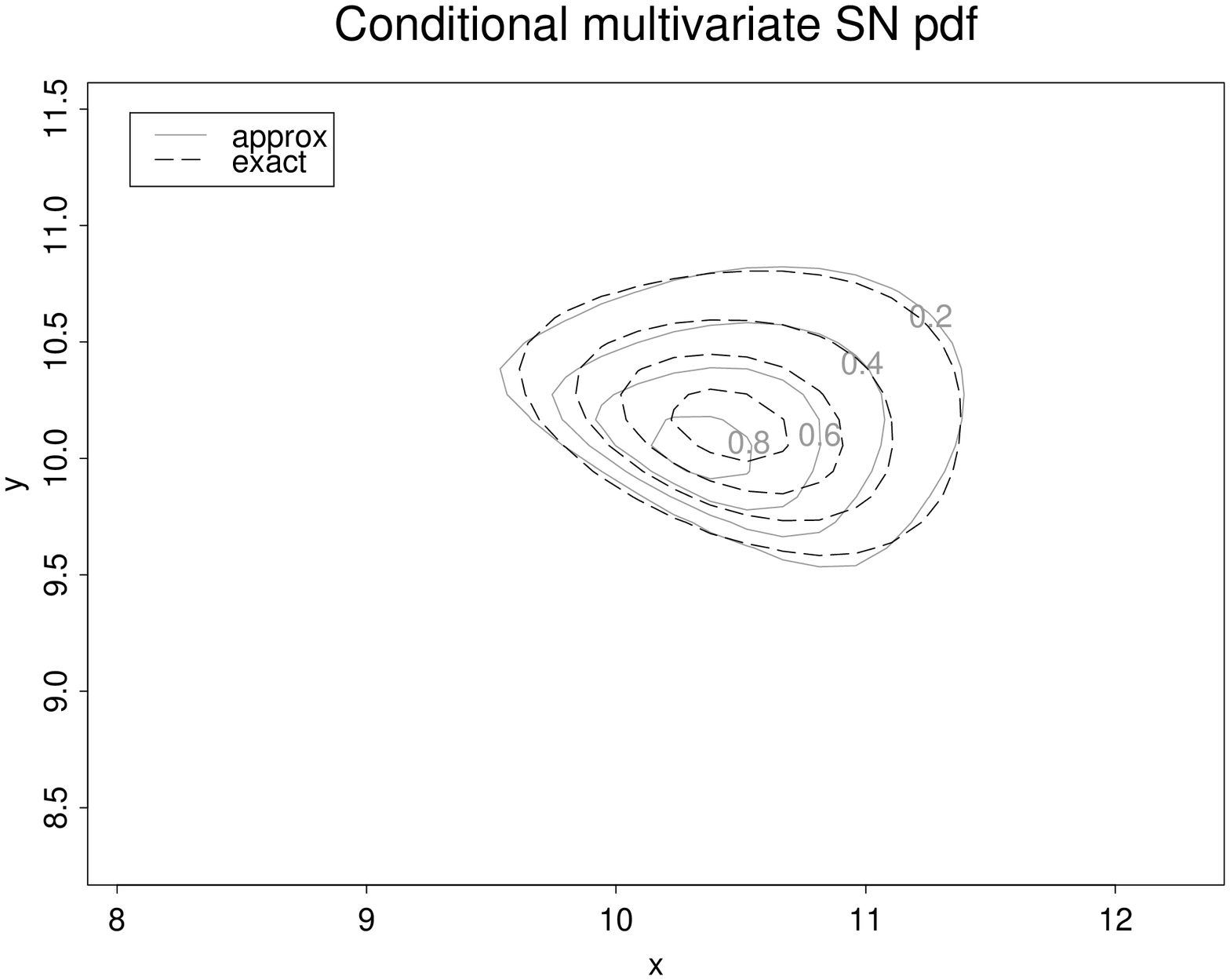}
  \hspace{1ex}
  \includegraphics[width=0.49\hsize,height=9cm]{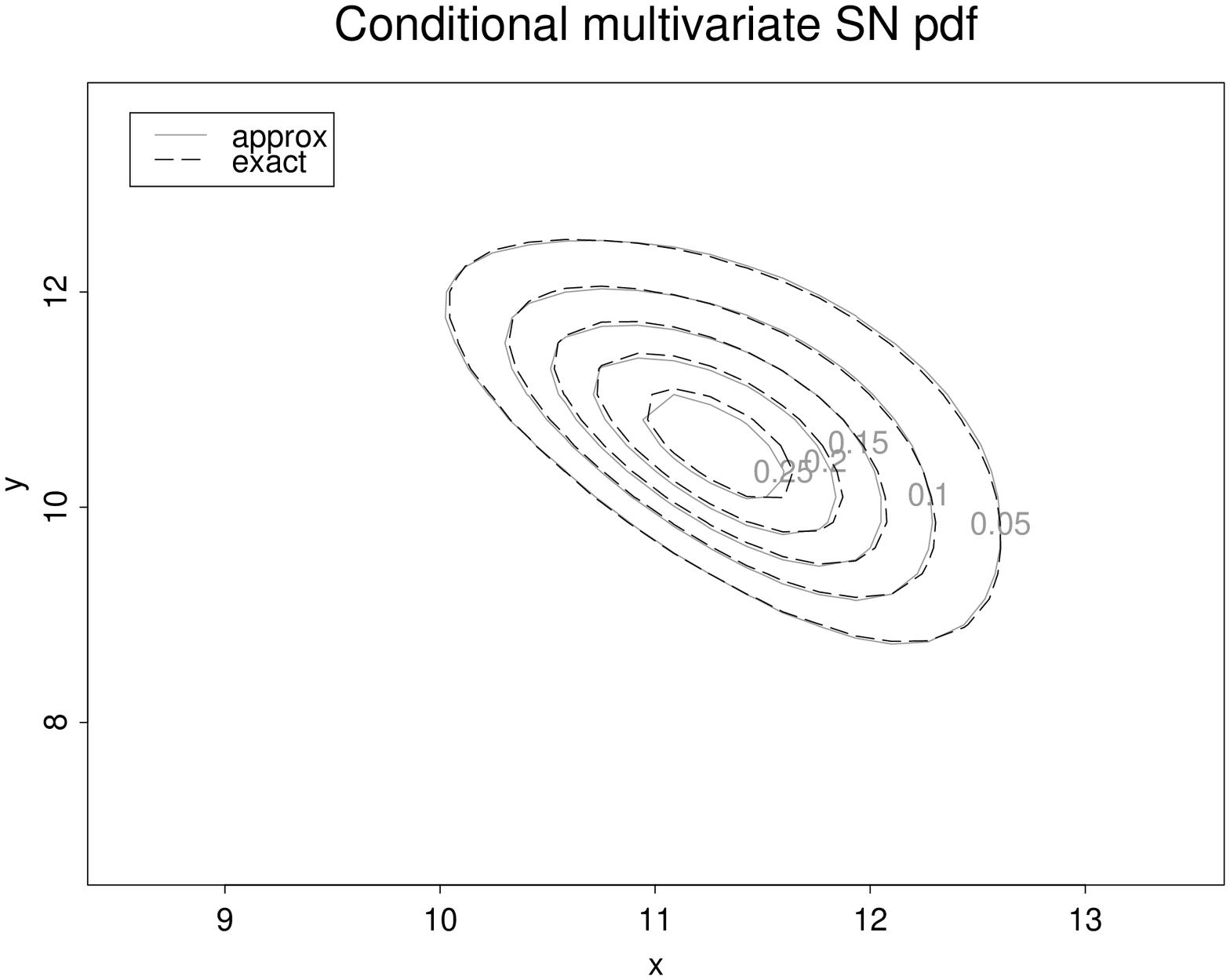}
  }
\caption{\small\sl Contour levels of the exact (dashed lines) and approximate
  (continuous lines) conditional density of a multivariate
  skew-normal variable, plotted for two sets of values of the parameters 
  and of the conditioning variable}
\label{fig:cond-pdf}
\end{figure}

The resulting equations allow explicit solution, except for extreme 
situations when the exact conditional density has an index of skewness
outside the range of the skew-normal one; these unfeasible cases are
very remote. 
In the overwhelming majority of cases, the equations can be solved, and
the approximate density is  close to the exact one.
Figure~\ref{fig:cond-pdf} shows the contour levels of the two
densities for two combinations of parameter values when $k-h=2$; 
the left panel shows one of the worst cases which have been observed,
while the right panel displays a much better, and also more
frequently observed, situation. 

Besides the generally small numerical discrepancy between the
approximate and the exact density, the following two properties hold.
\begin{itemize}
\item \emph{Independence is respected.}
If two components of $Y_2$ are independent conditionally on $Y_1=y_1$
with respect to the exact conditional density, so they are with
respect to the approximate one, and \emph{vice versa}.
\item \emph{Interchange of marginalization and conditioning.}
Integrating out some components of $Y_2$ after conditioning produces
the same result of integration followed by conditioning.  This fact is
obvious when using the exact density; it still holds for the
approximate one.
\end{itemize}
To prove the first statement, denote by $(a,b)$ a partition of  set of
indices composing $Y_2$.  Conditional independence of $Y_a$ and $Y_b$ 
implies that $\Omega_{22\cdot1}$ is block diagonal and that
one of the two components, $Y_a$ say, has no skewness; 
hence $\tilde\delta_a=0$ and $\tau_a=0$. Therefore all off-diagonal
blocks composing the variance in (\ref{f:Evar-cond}) are 0, and the same
structure must hold in the matching quantity of the approximating
distribution. The converse statement can be proved similarly.
To prove the second statement, simply notice that the approximation
preserves exact cumulants up to the third order, which uniquely
identify a member of the SN family;  hence also the cumulants of the
marginal distribution are preserved up to the same order.

The degree of accuracy of the approximation jointly with the above two
properties  support routine use of the approximate conditional
density in place of the exact one. 
In this sense, we can say that  the skew-normal class
of  density is closed with respect to the conditioning operation.

%==============================================================================

\section{Statistical issues in the scalar case}    \label{s:stat}

\subsection{Direct parameters}                     \label{s:direct}
Starting from this section, we switch attention to inferential
aspects, and other issues of more direct statistical relevance,
initially by considering univariate distributions.

Some of the issues discussed in this subsection have a close
connection with the problem considered by Copas \& Li (1997) 
and the sociological literature on Heckman's model referenced there; 
see also Aigner \emph{et al.} (1977) and the literature of stochastic
frontier models.

In the univariate case, write $Y\sim \SN(\xi,\omega^2,\alpha)$,
dropping the subscript $k$ for simplicity. If a random sample 
$y=(y_1,\ldots,y_n)\T$ is available, the loglikelihood function 
for the direct parameters $DP=(\xi,\omega, \alpha)$ is
\begin{equation}
  \ell(DP) = -n\log\omega  - \half z\T z 
                             + \sum_i \zeta_0(\alpha z_i)
  \label{f:logL-dp}
\end{equation}
where  $z= \omega\inv(y-\xi 1_n)$
% \begin{equation}
%                 z= \omega\inv(y-\xi)     \label{f:z}
% \end{equation}
and $z_i$ denotes its $i$-th component; here $1_n$ is the $n\times1$
vector of all ones.
We shall denote by $\hat\alpha$ the maximum likelihood estimate (MLE)
of $\alpha$, and similarly for the other parameters.
The likelihood equations are immediately written down, namely
\begin{eqnarray*}
 &&  \sum z_i - \alpha \sum p_{1i} = 0 , \\
 &&  \sum z_i^2 -  \alpha \sum p_{1i} z_i - n = 0, \\
 &&  \sum p_{1i} z_i =0
\end{eqnarray*}
where $p_{1i} = \zeta_1(\alpha z_i)$.
There are however two sort of problems with this parametrization. 
Firstly, there is always an inflection point at $\alpha=0$ of the 
profile loglikelihood. % $\ell^*(\alpha)$.  
Correspondingly, at  $\alpha=0$,  the expected Fisher information 
becomes singular.
This phenomenon is a special case of the problem studied in greater
generality by Rotnitzky \emph{et al.} (1999).

\begin{figure}
\psfrag{lambda}[][]{$\alpha$}
\psfrag{omega}[][]{$\omega$}
\psfrag{dataset: otis}{}
\psfrag{otis}{}
\psfrag{profile 2(log-Likelihood)}{}
\psfrag{profile relative 2(log-Likelihood)}{}
\psfrag{Profile relative 2(logLikelihood)}{}
\centerline{\includegraphics[width=0.99\hsize,height=9cm]{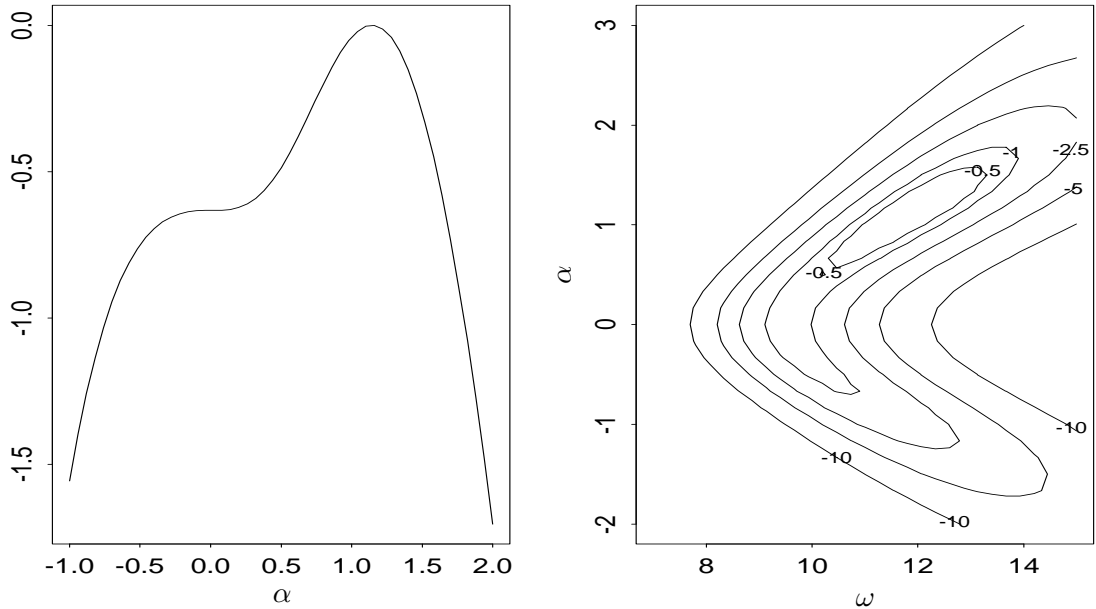}}
\caption{\small\sl Twice relative profile loglikelihood of $\alpha$ (left) and 
      contour levels of the similar function of
      $(\omega,\alpha)$ (right) for the Otis data,
      when the direct parametrization is used}
\label{fig:otis-dp}
\end{figure}

In addition,  the likelihood function itself can be problematic;
its shape can be far from quadratic even when $\alpha$ is not near 0.
This aspect is clearly illustrated by the plots given by
Arnold \emph{et al.}~(1993) who have analysed a dataset of size 87, 
later referred to as the Otis data;
see also Figure~\ref{fig:otis-dp}, which refers to the same data.

% From the shape of Figure~\ref{fig:otis-dp}, it is predicatable that direct
% maximization of $\ell(DP)$ can be problematic, and it fact  this is
% confirmed by various numerical examples not reported here.

For evaluation of the MLE, gradient-based methods have been considered,
but better results were obtained using the  EM algorithm, with the
introduction of a fictitious unobserved variable which 
is essentially $|X_0|$ of Proposition~\ref{th:stoch-repr}.
This method works satisfactorily, at least when the initial values are
chosen by the method of moments.
As typical for the EM algorithm, reliability rather than speed is its
best feature.  
Methods for accelerating the EM algorithm are available; see for
instance Meng \& van Dyk (1997) and references therein.
However, we prefer to expand in greater detail the discussion of
another approach, for the reasons explained in the next subsection.
 
\subsection{Centred parameters}                         \label{s:centred}
To avoid the singularity problem of the information matrix at $\alpha=0$,
Azzalini (1985) has reparameterized the problem by writing
\[    Y       = \mu+\sigma Z^\circ, \] 
where
\[   Z^\circ = (Z-\mu_z)/\sigma_z, \quad \sigma_z=\radice{1-\mu_z^2}, \]
and considering the centred parameters $CP=(\mu, \sigma,\gamma_1)$
instead of the DP parameters.  
Here $\gamma_1$ is the usual univariate index of skewness, which is equal to
the square root of the multivariate index of skewness of Section~\ref{s:cum}, 
taken with the same sign of $\alpha$.
Clearly, there is the correspondence
\[  
   \xi = \mu - \sigma\sigma_z\inv \mu_z, \qquad \omega= \sigma\sigma_z\inv.
\]

In the case of a regression problem, write $\E{Y_i}=x_i\T\beta$, where
$x_i$ is a vector of $p$ covariates and $\beta$ is vector parameter.
The corresponding loglikelihood is then
\[  
    \ell(CP) = n\log(\sigma_z/\sigma)-\half z\T z + \sum\zeta_0(\alpha z_i)
\]
where
\[  
     z_i = \mu_z+\sigma_z\sigma\inv(y_i-x_i\T\beta)= \mu_z+\sigma_z r_i, 
            \qquad
     z= (z_1,\ldots,z_n)\T.
 \]
In case we wanted to reformulate the regression problem in terms of direct
parameters, then only the first component must be adjusted, namely
\[  \beta_1^{DP} = \beta_1^{CP}-\sigma \mu_z/\sigma_z \]
in a self-explanatory notation.

\begin{figure}
\psfrag{gamma1}[][]{$\gamma_1$}
\psfrag{sigma}[][]{$\sigma$}
\psfrag{lambda}[][]{$\alpha$}
\psfrag{omega}[][]{$\omega$}
\psfrag{dataset: otis }{}
\psfrag{otis}{}
\psfrag{profile relative 2(logLikelihood)}{}
\psfrag{Profile relative 2(logLikelihood)}{}
\centerline{\includegraphics[width=0.99\hsize,height=9cm]{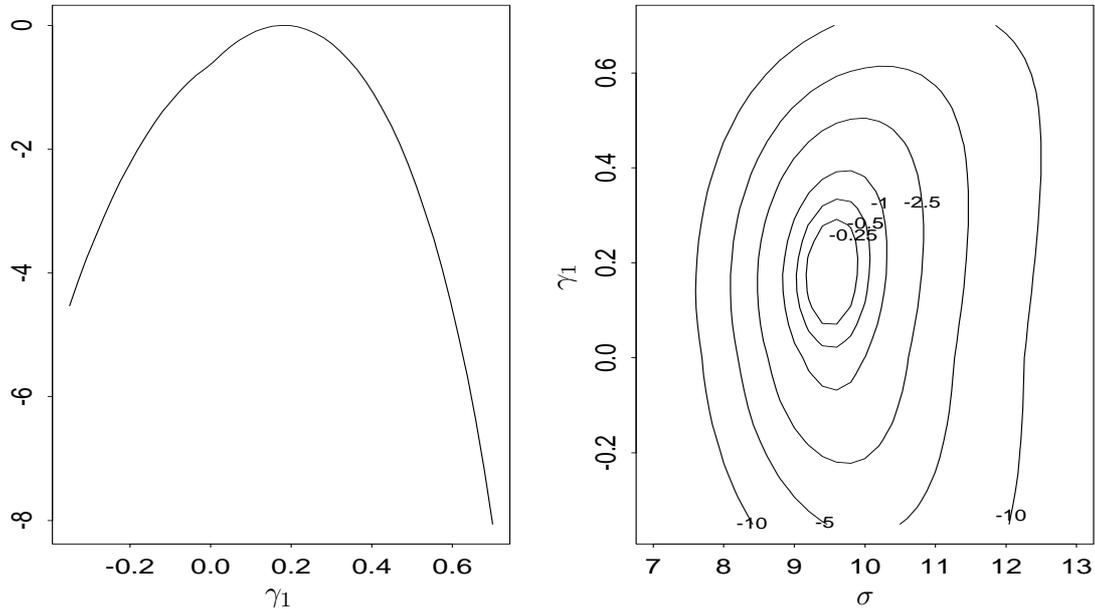}}
\caption{\small\sl Twice relative profile loglikelihood of $\gamma_1$ (left) and
    contour level of the similar function of 
      $(\sigma,\gamma_1)$ (right) for the Otis data,
      when the centred parametrization is used}
\label{fig:otis-cp}
\end{figure}

The gradient and the Hessian matrix of the loglikelihood in the CP
parametrization are more involved than with the DP parametrization, and
we confine the details in an appendix. 
The effects of the reparametrization are however beneficial in
various respects and worth the algebraic complications, for the
following reasons.
\begin{itemize}
\item 
The reparametrization removes the singularity of the information
matrix  at $\alpha=0$.
This fact was examined numerically by Azzalini (1985), and checked by 
detailed analytic computations by Chiogna (1997).
\item 
Although  not orthogonal, the components of CP are less correlated
than those of DP, especially  $\mu$ and the $\gamma_1$.
This fact can be checked numerically with the aid of the
expressions given in an appendix.
\item
The likelihood shape is generally much improved.  This is illustrated 
by Figure~\ref{fig:otis-cp}, which refers to the same data of 
Figure~\ref{fig:otis-dp};
the left panel refers to twice the relative profile
loglikelihood for the new shape parameter $\gamma_1$, and the right
panel refers to the pair $(\sigma,\gamma_1)$.  
There is a distinct improvement 
over the earlier figure, in various respects: 
\begin{itemize}
\item 
    the inflection point at $\alpha=0$ of the first panel of 
    Figure~\ref{fig:otis-dp} has been removed, with only a mild change of
    slope at $\gamma_1=0$ left;
\item 
    the overall shape of the profile loglikelihood has changed into one
    appreciably closer to a quadratic shape;
\item 
    near the MLE point, the axes of the approximating ellipsis are now
    more nearly alligned to the orthogonal axes than before.
\end{itemize}
\item
Simulation work, whose details are not reported here, showed that the
marginal distribution of $\hat\xi$ can be bimodal when $n$ and 
$|\alpha|$ are small or moderate; for instance it happens with $n=50$ ,
sampling from $\SN(0,\,1,\,1)$. Such an unusual  distribution of the MLE
is in qualitative agreement with the findings of Rotnitzky \emph{et
al.} (1999). Again, this unpleasant feature disappeared with the CP
parametrization, in the sense that the distribution of the new 
location parameter $\hat\mu$ exhibited a perfectly regular behaviour.
\end{itemize}

The advantages of CP over DP are not only on the theoretical side but
also practical,  since the more
regular shape of the loglikelihood leads to faster convergence of
the numerical maximization procedures when computing the MLE.

For numerical computation of the MLE, we have obtained satisfactory
results by adopting the following scheme:
(i) choose initial values by the method of moments;
(ii) optionally, improve these estimates by a few EM iterations;
(iii) obtain the MLE either by Newton--Raphson or by quasi-Newton methods.
Only in a few cases, the third stage did not  converge;
full EM iteration was then used, and this always led to convegence.

The set of S-Plus routines developed for these computations, as well
as those related to the problems discussed later, will be made freely
available on the WorldWideWeb.

\subsection{Anomalies of MLE}  \label{s:mle-weird}

Notwithstanding what is stated near the end of the previous
subsection, there are still cases where the likelihood shape and 
the MLE are problematic.
We are not referring here to difficulties with  numerical
maximization, but to the intrinsic properties of the likelihood
function, not removable by change of parametrization.

\begin{figure}
\centerline{\includegraphics[width=0.70\hsize,height=9cm]{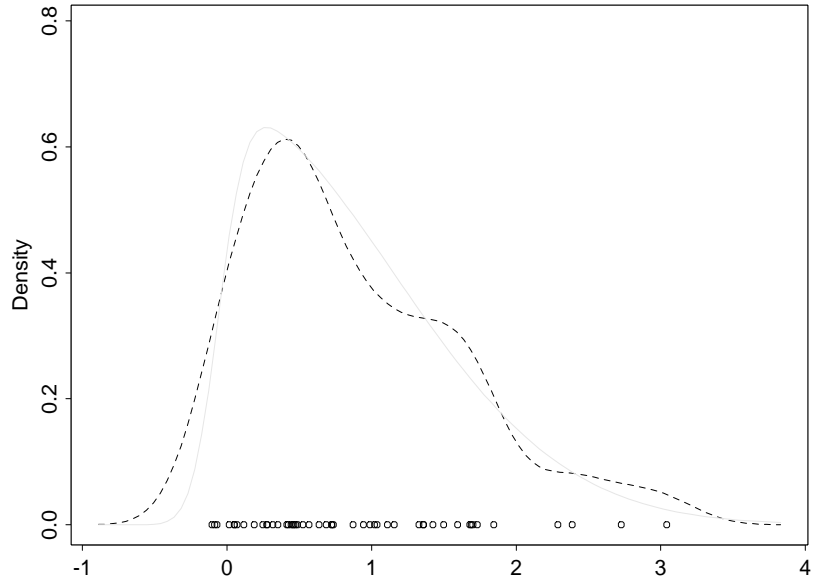}}
\caption{\small\sl Simulated data points (small circles) leading to
  $\hat\alpha=\infty$, with
  nonparametric density estimate (dashed curve) and parametric curve 
  with $\alpha=8.14$ (continuous curve)}
\label{fig:frontier}
\end{figure}

An illustration is provided by Figure~\ref{fig:frontier}; 
here 50 data points, sampled from $\SN(0,1,5)$, are plotted on the 
horizontal axis, together with a nonparametric estimate of the density
(dashed curve) and another (continuous) curve 
representing a skew-normal density. This parametric curve has 
$\alpha=8.14$ but it is not the one of the MLE, however: 
the MLE has $\alpha=\infty$, which corresponds to the half-normal density.

This divergence of $\hat\alpha$  (or equivalently
$\hat\gamma_1\to 0.99527$, its maximal value) looks rather surprising,
since apparently there is nothing pathological in the data
pattern of Figure~\ref{fig:frontier}; the sample index of skewness
is 0.9022, which is inside the feasible region of $\gamma_1$.
Similar situations occur with a non-negligible frequency when $n$ is
small to moderate, but they disappear when $n$ increases.

The source of this sort of anomaly is easy to understand in the one-parameter
case with  $\xi$ and $\omega$ known;  $\xi=0$, $\omega=1$, say.
If all sample values have the same sign, the final term of
(\ref{f:logL-dp}) increases with $\pm\alpha$, depending the sign 
of the data but irrespective of their actual values,
as it has been remarked by Liseo (1990).
For instance, if 25 data are sampled from $\SN(0,1,5)$, the
probability that they are all positive is about 0.20.

When all three  DP parameters are being estimated, the explanation of 
this fact is not so clear, but it is conceivable that a similar
mechanism is in action.

In cases of this sort, the behaviour of the MLE appears qualitatively
unsatisfactory, and an alternative estimation method is called for.
Tackling this problem is beyond the scope of the present paper, however.
As a temporary solution we adopted the following simple strategy:
when the maximum occurs on the frontier, re-start the maximization 
procedure  and stop it when it reaches a loglikelihood value
not significantly lower than the maximum.  
This was the criterion used for choosing the  parametric 
curve plotted in Figure~\ref{fig:frontier}; in this case the
difference from the maximum of the loglikelihood is 2.39, far below
the 95\% significant point of a $\chi^2_3/2$ distribution.

The above proposal leaves some degree of arbitrariness, 
since it does not say exactly how much below the maximum to stay. 
In practice the choice is not so dramatic, because the boundary 
effect involves only $\alpha$, and when this is large,
$\alpha>20$ say, the actual shape of the density varies very slowly.
Moreover, in the numerical cases  which have been examined,
the loglikelihood function was very flat only along the $\alpha$-axis,
while it was far more curved with along the location and scale 
parameters which were then little affected by the 
specific choice of $\alpha$, within quite wide limits.

%==============================================================================

\section{Applications to multivariate analysis}    \label{s:multivar}
%---------------------
\subsection{Fitting multivariate distributions}    \label{s:fit.msn}

In the case of independent observations $(y_1,\ldots,y_n)$
sampled from  $\SN_k(\xi_i,\Omega,\alpha)$ for $i=1,\ldots,n$, 
the loglikelihood is
\begin{equation}
     \ell = -\half n \log|\Omega|
                  -\half n\, \tr(\Omega\inv V)
                  + \sum_i \zeta_0\{\alpha\T \omega\inv(y_i-\xi_i)\}
                                              \label{f:logL}
\end{equation}
where
\[     % z_i = \omega\inv(y_i-\xi_i), \qquad    
       V   = n\inv \sum_i (y_i-\xi_i) (y_i-\xi_i)\T.
\]
The location parameters have been considered to be different having in
mind a regression context where $\xi_i$ is related to $p$
explanatory variables $x_i$ via
\[   \xi_i\T = x_i \beta,   \qquad (i=1,\ldots,n), \]
for some $p\times k$ matrix $\beta$ of parameters.

It would be ideal to reproduce in this setting the centred
parametrization introduced in the scalar case. 
This approach poses difficulties, and we follow  a different 
direction to obtain the MLE. Once the estimates have been computed, they
could be converted componentwise to the centred parameters.

The letters $y$, $X$, $\xi$ will denote the matrices 
of size $n\times k$, $n\times p$ and $n\times k$ containing
the $y_i$'s, the $x_i$'s, and the $\xi_i$'s, respectively.
Also, a notation of type $\zeta_m(z)$ represents the vector obtained by
applying the function $\zeta_m(\cdot)$ to each element of the vector $z$.

Regarding  $\eta=\omega\inv\alpha$ as a parameter in replacement of
$\alpha$ separates the parameters in (\ref{f:logL}) in the following sense: 
for fixed $\beta$ and $\eta$,
maximization of $\ell$ with respect $\Omega$ is equivalent to maximizing
the analogous function for normal variates for fixed $\beta$, which has
the well known solution
\[   \hat\Omega(\beta) = V(\beta) = n\inv u\T u \]
where $u= (y-X\beta)$.
Replacing this expression in $\ell$ gives the profile loglikelihood
\[ 
   \ell^*(\beta,\eta) = 
        -\half n\log|V(\beta)| -\half nk + 1_n\T \zeta_0(u\eta)
\]
with substantial reduction of dimensionality of the maximization
problem. Numerical maximization of $\ell^*$ is required; this process
can be speeded up substantially if the partial derivatives
\begin{eqnarray*}
  \pd{\ell^*}{\beta} &=& X\T u\, V(\beta)\inv - X\T\zeta_1(u\eta\,)\eta\T, \\
  \pd{\ell^*}{\eta}  &=& u\T \zeta_1(u\eta), 
\end{eqnarray*}
are supplied to a quasi-Newton algorithm.
Upon convergence, numerical differentiation of the gradient leads to
approximate standard errors for $\beta$ and $\eta$, hence  for
$\alpha$ after multiplication by $\omega$.

\begin{figure}
\centerline{\includegraphics[width=\hsize,height=12cm]{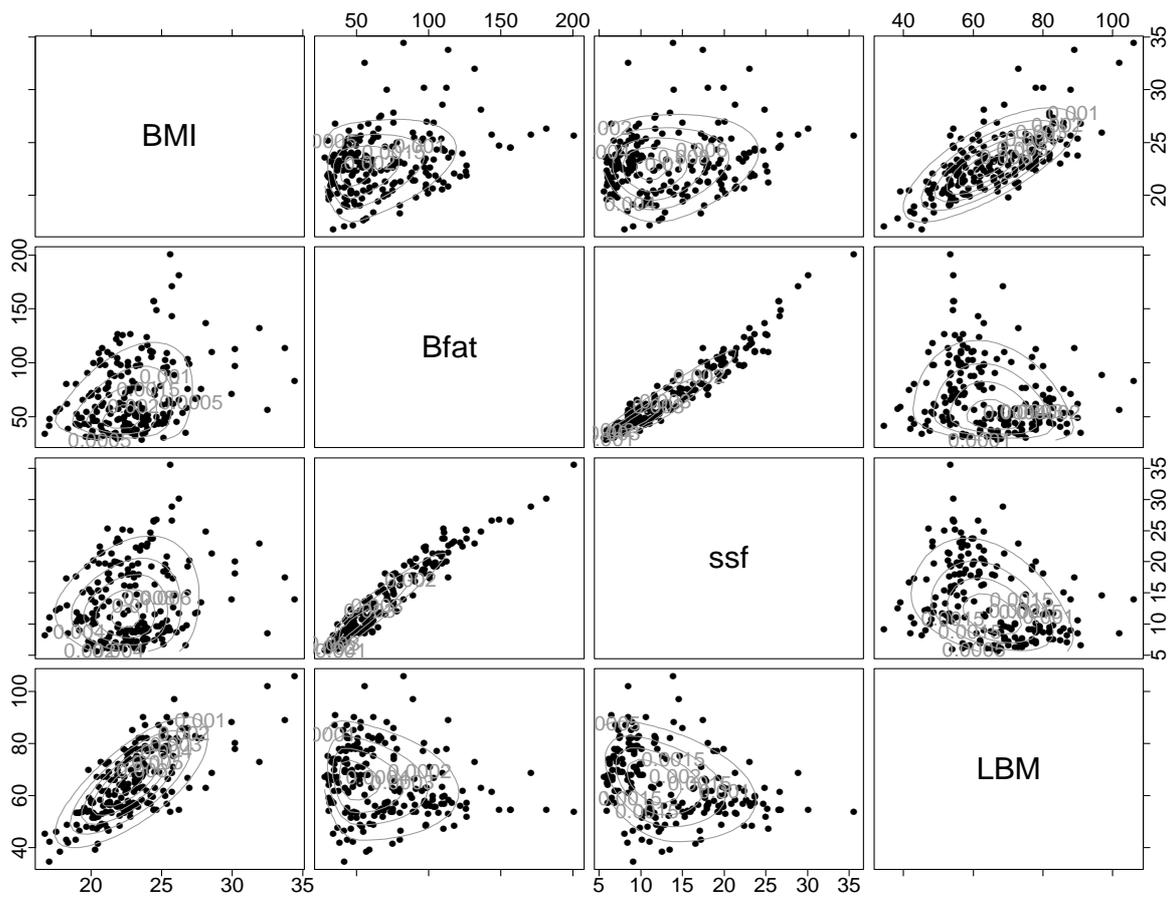}}
\caption{\small\sl Scatterplots of some pairs of the AIS variables and contour
         levels of the fitted distribution}
\label{fig:ais-pair}
\end{figure}

The above computational scheme has been used satisfactorily in
numerical work with non-trivial dimensions of the arrays $X$, $y$, $\beta$. 
A very simple illustration is provided by Figure~\ref{fig:ais-pair}
which refers to a subset of the AIS (Australian Institute of Sport) data 
examined by Cook \& Weisberg (1994), 
which contains various biomedical measurements on a group of 
Australian athletes; we then have $k=4$, $p=1$, $n=202$.
Figure~\ref{fig:ais-pair} displays  the scatter plot of each pair of 
the four variables considered superimposed with the contour lines of the
marginal density obtained by marginalization of the fitted $\SN_4$
density. 

Visual inspection of Figure~\ref{fig:ais-pair} indicates a
satisfactory fit of the density to the data.
However, to obtain a somewhat more comprehensive graphical display,
consider the Mahalanobis distances
\begin{equation}
   d_i = (y_i-\xi)\T \Omega\inv (y_i - \xi), \qquad (i=1,\ldots,n),
   \label{f:mahal}
\end{equation}
which are sampled from a $\chi^2_k$ if the fitted model is
appropriate, by using Proposition~\ref{quad-dist}.
In practice, estimates must be replaced to the exact
parameter values in (\ref{f:mahal}).
The above $d_i$'s must be sorted and plotted versus the $\chi^2_k$
percentage points.  Equivalently, the cumulative $\chi^2_k$
probabilities can be plotted against their nominal values 
$1/n,2/n,\ldots,1$; the points should lie on the bisection line of the
quadrant. 
This diagnostic method is a natural analogue of a well-know 
diagnostics used in normal theory context (Healy, 1968).

\begin{figure}
\centerline{\includegraphics[width=0.99\hsize,height=9cm]{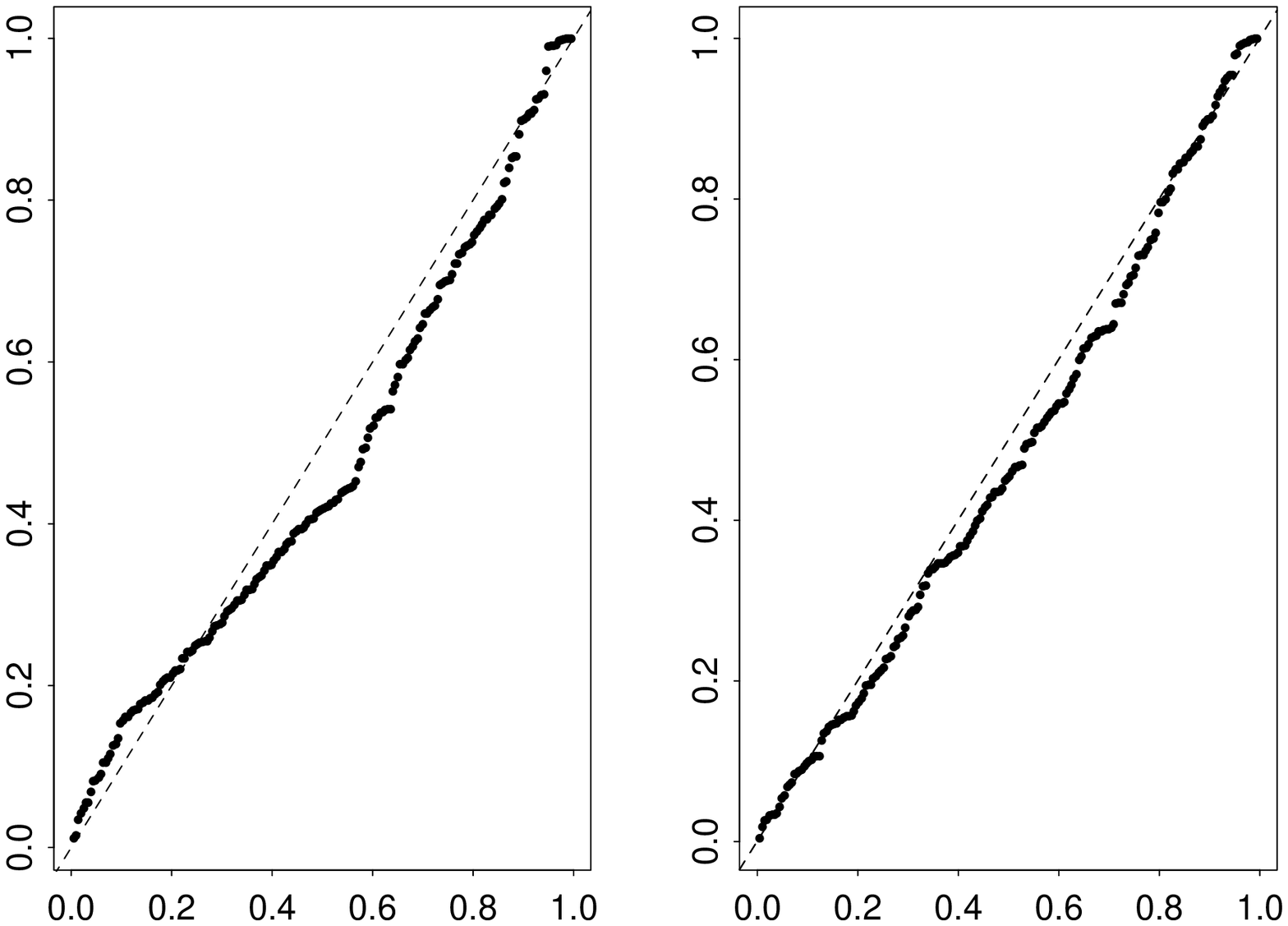}}
\caption{\small\sl Healy's plot when either a normal distribution (left panel) 
     or a skew-normal distribution (right panel) is fitted to the AIS data}
\label{fig:ais-pp}
\end{figure}

Figure~\ref{fig:ais-pp} diplays the second variant of this plot  for
the AIS data, in its right-hand side panel; 
the left-hand side panel shows the similar traditional plot under
assumption of normality. 
Comparison of the two plots indicates a substantial improvement of the
skew-normal fit over the normal one. 

A similar conclusion is achieved by considering a parametric test for 
normality  which is provided by the likelihood ratio test for the null
hypothesis $\alpha=0$, that is
\[   
   2\{ \ell(\hat\xi,\hat\Omega,\hat\alpha)- \ell(\hat\mu, \hat\Sigma,0) \}
\]
where $(\hat\mu, \hat\Sigma)$ denote the MLE of $(\xi, \Omega)$
under the assumption of normality.
The observed value of the test statistics in the above example is
over 103, and the associated
value of the $\chi^2_4$ distribution function does not even need 
to be computed.

%---------------------
\subsection{Discriminant analysis}                 \label{s:discrim}

The results of Section~\ref{s:transforms}, once reinterpreted in the
more general setting introduced in Section~\ref{s:3param}, provide 
tools to examine the behaviour of  many classical multivariate
techniques, when based on linear transforms of the data, in the more
general context of $\SN$ variables. 
For  the present discussion, however, we shall restrict ourselves to
a rather simple problem of discrimination between two populations,
under the traditional hypothesis that they differ only in the location
parameters.

If $Y_i\sim \SN_k(\xi_i, \Omega, \alpha)$ denote the random variables
associated to the two populations ($i=1,2$), then the likelihood-based
discrimination rule allocates a new unit with observed vector $y$ to
population 1 if
\begin{equation}     \label{f:discr-sn} 
        (\xi_1 - \xi_2)\T \Omega\inv (y - \half(\xi_1 + \xi_2))
                   +\zeta_0(w_1)-\zeta_0(w_2) + \log(\pi_1/\pi_2) > 0 
\end{equation}
where $w_i=w_i(y)=\alpha\T\omega\inv(y-\xi_i)$ and $\pi_i$ is the prior
probability of the $i$-th  population  ($i=1,2$).

Nonlinearity of the left-hand side  of the above inequality
prevents explicit solution. 
However, some properties can be obtained; 
% Departure from linearity of the above function involves consideration
% of  its gradient, which depends on
% \[          \zeta_1(w_1(y))-\zeta_1(w_2(y)). \]  
one is that the likelihood-based discriminant function is
a linear function of $y$  when either of the following  conditions
holds:
\begin{eqnarray}    
      (\xi_1 - \xi_2)\T \omega\inv \alpha  = 0,    \label{f:cond-lin1} \\
      \omega\inv\alpha = c\,\Omega\inv(\xi_1-\xi_2) \label{f:cond-lin2}
\end{eqnarray}
where $c$ ia non-zero scalar constant. The proof is omitted.

The natural alternative to (\ref{f:discr-sn}) is the Fisher linear 
discriminant functions, whose commonly used expression is
\[
    (\mu_1 - \mu_2)\T \Sigma\inv \left(y- \half(\mu_1 + \mu_2) \right)
      + \log(\pi_1/\pi_2)
    > 0,
\]
using a self-explanatory notation; in the present case, this can be
re-written as
\begin{equation}  \label{f:Fisher-sn}
     (\xi_1 - \xi_2)\T (\Omega - \omega \mu_z \mu_z\T \omega)\inv 
    \left( y- \half(\xi_1 + \xi_2 + 2 \omega \mu_z) \right) 
    + \log(\pi_1/\pi_2)>0.
\end{equation}

\begin{proposition}
When condition (\ref{f:cond-lin1}) holds, the discriminant rules 
(\ref{f:discr-sn}) and (\ref{f:Fisher-sn}) coincide.
\end{proposition}
\emph{Proof.} First, notice that (\ref{f:cond-lin1}) implies
that $w_1(y)=w_2(y)$ in (\ref{f:discr-sn}). Next, use 
(\ref{f:smw}) to invert  
$(\Omega - \omega \mu_z \mu_z\T \omega)$ in (\ref{f:Fisher-sn}),
leading to
\[
   (\xi_1 - \xi_2)\T \Omega\inv (y-\half(\xi_1+\xi_2)- \omega\mu_z) > 0.
\]
Then, on using  (\ref{f:cond-lin1}) again and  
noticing that  the vectors  $\Omega\inv \omega \mu_z$ and  
$\omega\inv \alpha$ have the  same direction, one obtains the result.
\par\vskip 2ex

In the general case,  (\ref{f:discr-sn}) and  (\ref{f:Fisher-sn})  
can only be compared numerically. 
The various cases considered differ for the 
relative positions of the locations parameters, while the other
parameters have been kept fixed; specifically, we have set 
$k=2$, $\pi_1=\pi_2$,  $\omega=I_2$,
$\Omega$  equal to the correlation matrix with off-diagonal elements 
equal to 0.4, $\alpha=(3,3)\T$, and $\|\xi_1-\xi_2\|^2=1$.
This choice of the parameters, such that  $\alpha$ is an eigenvector
of $\Omega$, has been made for the sake of simplicity, in the
following sense.
It turns out that the quantities regulating the basic behaviour of the
classification rules are the angle  $\theta_1$ between the vectors
$\omega\inv\alpha$ and $\xi_1-\xi_2$, and the angle $\theta_2$ between
$\omega\inv\alpha$ and $\Omega\inv(\xi_1-\xi_2)$.
The above choice of $\alpha$ and $\Omega$ makes it easier to choose 
values of $\xi_1-\xi_2$ fulfilling conditions (\ref{f:cond-lin1}) and
(\ref{f:cond-lin2}), i.e. such that $\cos \theta_1=0$ and 
$\cos \theta_2=1$. 

\begin{figure}
\centerline{\includegraphics[width=0.99\hsize,height=13cm]{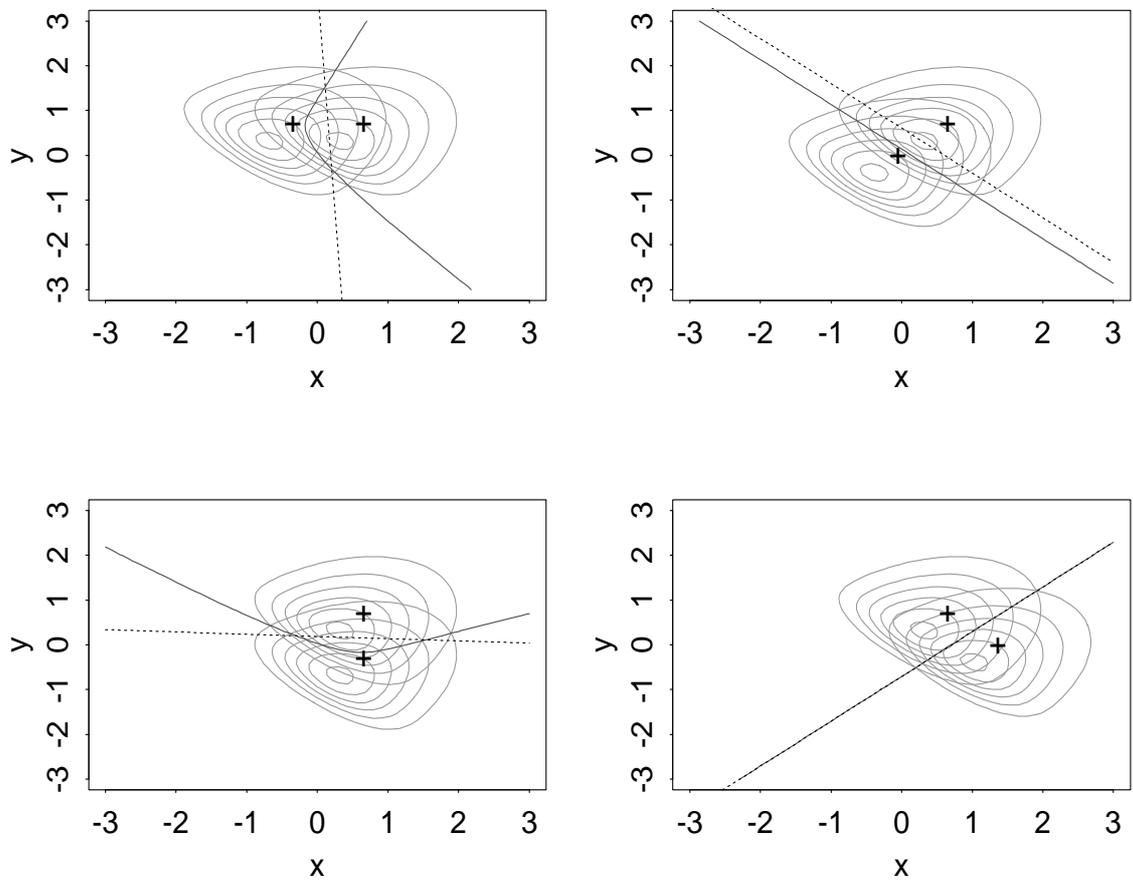}}
\caption{\small\sl Contour plots of four pairs of $\SN_2$ variables, with
  likelihood discriminant function (continuous line) and Fisher
  linear discriminant function (dashed line)
  }
\label{fig:discr}
\end{figure}

Figure~\ref{fig:discr}  shows the relevant entities for a few cases.
Each panel of the figure displays the contour levels of the two 
population densities with superimposed the separation lines of 
the two  discriminant rules.
The bottom-right panel corresponds to a case satisfying
(\ref{f:cond-lin1}) and only one  discrimination line is then visible;
the top-right panel corresponds to fulfilling (\ref{f:cond-lin2})
and the two discriminant lines are parallel.

Table~\ref{t:simul} contains summary values of the numerical work, in
particular misclassification probabilities, for a larger number of cases.
For the Fisher rule,  classification probabilities can be computed 
exactly with the aid of (\ref{f:lin-transf});
for (\ref{f:discr-sn}), the corresponding probabilities have been 
evaluated by simulation methods, using 100000 replicates for each case.
The main qualitative conclusions from these figures are as follows:
(a) the total misclassification probability is lower for the
likelihood-based rule than for the Fisher linear discriminant, 
as expected from known results (Rao, 1947);
(b) the Fisher rule is however not much worse than the other one, and
its two components are more balanced that the analogous ones of the
likelihood-based rule, which could be considered as an advantage on
its own;
(c) for some values of $\theta_1$ and $\theta_2$, the fraction of
cases which are classified differently by the two rules is not
negligible; hence the choice of the method can be relevant
even if the probabilities of misclassification are similar.

\begin{table}
\begin{center}
\begin{tabular}{|rr|rr|r|rr|}
\hline
$ p_{1L}$ &  $p_{1F}$ & $ p_{2L}$ &  $p_{2F}$ 
               & $p^*$ & $\cos \theta_1$ & $\cos \theta_2$\\
\hline
0.35 & 0.23 & 0.10 & 0.28 & 0.84 &  1.000 &  1.000 \\
0.35 & 0.23 & 0.11 & 0.28 & 0.85 &  0.907 &  0.981 \\
0.34 & 0.23 & 0.13 & 0.27 & 0.87 &  0.719 &  0.924 \\
0.31 & 0.23 & 0.16 & 0.26 & 0.89 &  0.530 &  0.831 \\
0.29 & 0.24 & 0.19 & 0.26 & 0.91 &  0.394 &  0.707 \\
0.27 & 0.25 & 0.21 & 0.26 & 0.92 &  0.275 &  0.556 \\
0.26 & 0.26 & 0.24 & 0.26 & 0.94 &  0.175 &  0.383 \\
0.26 & 0.26 & 0.25 & 0.26 & 0.96 &  0.085 &  0.195 \\
0.26 & 0.26 & 0.26 & 0.26 & 1.00 &  0.000 &  0.000 \\
0.25 & 0.26 & 0.26 & 0.26 & 0.96 & -0.085 & -0.195 \\
0.24 & 0.26 & 0.26 & 0.26 & 0.94 & -0.175 & -0.383 \\
0.21 & 0.26 & 0.27 & 0.25 & 0.92 & -0.275 & -0.556 \\
0.19 & 0.26 & 0.29 & 0.24 & 0.91 & -0.394 & -0.707 \\
0.16 & 0.26 & 0.31 & 0.23 & 0.89 & -0.530 & -0.831 \\
0.13 & 0.27 & 0.33 & 0.23 & 0.87 & -0.719 & -0.924 \\
0.10 & 0.28 & 0.35 & 0.23 & 0.85 & -0.907 & -0.981 \\
0.10 & 0.28 & 0.35 & 0.23 & 0.84 & -1.000 & -1.000 \\
%  0.29 & 0.24  &  0.19  & 0.26  & 0.91  &  0.394  &  0.707 \\
%  0.34 & 0.23  &  0.13  & 0.27  & 0.87  &  0.719  &  0.924 \\
%  0.35 & 0.23  &  0.10  & 0.28  & 0.84  &  1.000  &  1.000 \\
%  0.34 & 0.23  &  0.13  & 0.27  & 0.87  &  0.719  &  0.924 \\
%  0.29 & 0.24  &  0.19  & 0.26  & 0.91  &  0.394  &  0.707 \\
%  0.26 & 0.26  &  0.24  & 0.26  & 0.94  &  0.175  &  0.383 \\
%  0.26 & 0.26  &  0.26  & 0.26  & 1.00  &  0.000  &  0.000 \\
%  0.24 & 0.26  &  0.26  & 0.26  & 0.94  & -0.175  & -0.383 \\
%  0.19 & 0.26  &  0.29  & 0.24  & 0.91  & -0.394  & -0.707 \\
%  0.13 & 0.27  &  0.34  & 0.23  & 0.87  & -0.719  & -0.924 \\
%  0.10 & 0.28  &  0.36  & 0.23  & 0.85  & -1.000  & -1.000 \\
%  0.11 & 0.27  &  0.34  & 0.23  & 0.87  & -0.719  & -0.924 \\
%  0.19 & 0.26  &  0.29  & 0.24  & 0.91  & -0.394  & -0.707 \\
%  0.24 & 0.26  &  0.26  & 0.26  & 0.94  & -0.175  & -0.383 \\
%  0.26 & 0.26  &  0.26  & 0.26  & 1.00  &  0.000  &  0.000 \\
%  0.26 & 0.25  &  0.24  & 0.26  & 0.94  &  0.175  &  0.383 \\
\hline 
\end{tabular}
\end{center}
\caption{\small\sl Classification probabilities of likelihood-based and Fisher
linear discriminant rules. The entries are:
$p_{1L}$, misclassification error probability using 
       likelihood based rule, when sampling from population 1;
$p_{1F}$,  misclassification error probability using 
      Fisher linear discriminant function, when sampling from population 1;
$p_{2L}$ and $p_{2F}$ are similar quantities in the case of
       sampling from population 2;
$p^*$, probability that the discriminant rules coincide;
$\theta_1$ and $\theta_2$ are angles associated to the relative
   position of the location parameters, as described in the text}
\label{t:simul}
\end{table}

% We have applied the two rules to enzime data collected from 218 patients
% with liver diseases; see Albert \& Harris (1987, chapter~5).

For numerical illustration, we have applied the two discriminant rules
to the same subset of the AIS data used in subsection \ref{s:fit.msn}.
The individuals were divided by sex, obtaining two groups of 102
male and 100 female athletes, respectively, and prior probabilities
were set equal to the observed frequencies. In this case $\theta_1 = 1.54041$
radians, a situation not so far from the one associated with 
(\ref{f:cond-lin1}), i.e coincidence of the two discriminant functions.
In fact the total number of misclassified subjects differs only for one unit:
more precisely Fisher rule fails in three units, while the likelihood-based
one fails in two.
Further numerical work has been done using data reported by  
Albert \& Harris (1987, chapter~5), fairly often used for illustration
in the context of discriminant methods. An overall sample of 218 individuals 
affected by liver problems are divided into four groups, corresponding
to severity of their status:
acute viral hepatitis (group $G_1$, 57 patients), 
persistent chronic hepatitis ($G_2$, 44 patients), 
aggressive chronic hepatitis  ($G_3$, 40 patients), and 
post-necrotic cirrhosis ($G_4$, 77 patients).
Albert \& Harris~(1987)  construct a discrimination rule
based on data on three  of four available liver enzymes:
aspartate aminotransferase (AST), alanine aminotransferase (ALT) and 
glutamate dehydrogenase (GLDH); 
the data have been logarithmically transformed because of extreme 
skewness in the original variables.
To ease comparison, we employed the same variables and applied the same
data transformation.

Goodness-of-fit and graphical diagnostics, along the lines of
subsection~\ref{s:fit.msn},
confirm the adequacy of the skew-normal distribution in
modeling this set of variables. 
Prior probabilities were set equal to the observed frequencies, i.e.
$\pi_1=0.26$, $\pi_2=0.20$, $\pi_3=0.18$ and $\pi_4=0.35$.
The summary results, shown in Table~\ref{t:hepatic},
indicate a slight improvement using the SN distribution instead of the
normal one, in the sense that 3 data points which were incorrectly
classified by the Fisher rule are now correctly classified, and only
one is moved in the reverse direction.

\begin{table}
\begin{center}
\begin{tabular}{lcccc}
\hline
     &  \multicolumn{4}{c}{Actual groups}   \\
Allocated groups & $G_1$ & $G_2$ & $G_3$ & $G_4$\\
\hline
   $G_1$    & 55, 55 &  5, 5  &  2, 2  & 0, 0  \\
   $G_2$    & 2, 2   & 36, 37 &  2, 4  & 0, 0  \\
   $G_3$    & 0, 0   &  0, 0  & 22, 20 & 10, 11 \\
   $G_4$    & 0, 0   &  3, 2  & 14, 14 & 67, 66 \\
   \hline
Total    & 57 & 44 & 40 & 77 \\
\hline
\end{tabular}
\end{center}
\caption{\small\sl Discrimination of the four groups of the hepatic data;
  the data indicate the number of individuals classified by
  likelihood rule (first entry) and by the Fisher discriminat
  function (second entry) }
\label{t:hepatic}
\end{table}

%---------------------

\subsection{Regression and graphical models} \label{s:graph}

Graphical models are currently a much studied research topic.
This subsection examines some related issues when the 
assumption of normal distribution of the variable is replaced by 
(\ref{f:dens}). We adopt Cox \& Wermuth (1996) as a reference text
for background material.

In the construction of a graphical model of normal variables, 
a key ingredient is the concentration matrix, 
i.e. the inverse of the covariance matrix, 
possibly scaled to obtain unit diagonal elements.
When the $(i,j)$-th entry of the concentration matrix is 0, this
indicates that the two corresponding components, $Y_i$ and $Y_j$ say,
are independent conditionally on all the others.
The associated concentration graph has then no edge between $Y_i$ and 
$Y_j$.
 
The results of sections~\ref{s:transforms} and \ref{s:3param}
enable us to transfer the above scheme in the context of
skew-normality; consider in particular Proposition~\ref{th:lin2}
and expression (\ref{f:sncond}).
Hence, two components, $Y_i$ and $Y_j$ say, of
$Y \sim \SN_k (\xi, \Omega, \alpha)$ are independent conditionally
on the others if the $(i,j)$-th entry of $\Omega\inv$ is zero 
and at most one of $\alpha_i$ and $\alpha_j$ is different from zero.
Hence $\Omega\inv$ plays a role analogous to the concentration
matrix in normal theory context, but also $\alpha$ must be considered
now.

Building a graphical model from real data involves to follow essentially 
the strategy presented by Cox \& Wermuth (1996) for the normal case.
The main difference is in the distinction between regression and
conditioning, which are essentially coincident in the normal case but
not here.

Since it seems best to illustrate the actual construction of a
graphical model in a specific example, we consider the data analysed
by Cox \& Wermuth (1996, chapter~6), concerning 68 patients with fewer
than 25 years of diabetes.  This dataset is of rather small sample
size for an adequate fitting of a multivariate SN distribution, but it
has been adopted here because it is a `standard' one in this context.
For each patient, eight variables are recorded; of these, glucose
control ($Y$) and knowledge about illness ($X$) are the primary
response and the intermediate response variables, respectively; the
special role of these two variables drives the subsequent analysis.
Of the other variables, $W$, $A$ and $B$ are explanatory variables regarded
as given, with $A$ and $B$ binary; $Z$, $U$ and $V$ are other stochastic
variables.  See the above reference for a full description of the
variables and some background information.

A preliminary analysis, using the methods described at the end of 
subsection~\ref{s:fit.msn}, shows the presence of a significant 
skewness in the distribution of some of the variables; this is
largely due to the $X$ component but not only to this one.
Therefore, we introduce a multivariate regression model of type
\[     (Y,X,Z,U,V) \sim \SN_5(\xi, \Omega, \alpha)  \]
where $\xi$ is a linear combination of $(1,W,A,B)$, and $\Omega$ and
$\alpha$ are constant across individuals.
Fit of the above model, using the algorithm
described in section~\ref{s:fit.msn}, led to a boundary solution, in
the sense that the components of $\hat\alpha$ diverged. 
Adopting the simple method described in section~\ref{s:mle-weird} to
handle these cases, a set of parameters has been chosen inside the 
parameter space having a loglikelihood value about 7.7 units lower than 
the maximum, which is a very minor loss in consideration of the large 
number of parameters being estimated.

Table~\ref{t:conc-5} gives the partial correlation matrix, $\hat\Omega^*$,
which is  $\hat\Omega\inv$ after scaling  to obtain unit diagonal entries
and changing signs of the off-diagonal entries, 
and  the shape parameters with standard errors and $t$-ratios.

Because of the different role played by the variables in the present 
problem, the most relevant entries of
Table~\ref{t:conc-5} are those of the first two rows of $\Omega^*$.
Joint inspection of both components of Table~\ref{t:conc-5}
indicates conditional independence of $(Y,Z)$, $(Y,U)$ and $(Y,V)$, while
there is conditional dependence between $(X,Z)$ and between $(Y,X)$.
Moreover the results concerning the regression component 
suggest dropping $B$ from the model.

\begin{table}\centering
\vbox{
  {$\hat\Omega^*= \bordermatrix{
           &    Y    &  X     &  Z      &  U     &  V     \cr
        Y  &    1    & -0.49  & \M0.09  &  -0.16 &\M0.06  \cr
        X  &   -0.49 &     1  &  -0.38  &  -0.04 &\M0.17  \cr
        Z  &  \M0.09 &  -0.38 &     1   & \M0.42 & -0.25  \cr
        U  &   -0.16 &  -0.04 & \M0.42  &      1 & -0.07  \cr
        V  &   -0.06 & \M0.17 &  -0.25  &  -0.07 & \M1.00 }$
}\vskip 2ex
{
\begin{tabular}{l r r r r r }
             &    Y  &  X  &  Z  &  U  &  V  \\
\hline
$\hat\alpha$ & 1.53  &  -32.89  & -3.49  &  -1.16  &  -2.41 \\
std.error    &  6.4  &   11.68  &  2.89  &   7.27  &   2.70\\
$t$-ratio    & 0.24  &   -2.81  & -1.21  &  -0.16  &  -0.89 \\
\hline
\end{tabular}
}
}%vbox
\caption{\small\sl Matrix $\hat\Omega^*$, $\hat\alpha$ and other quantities 
         associated to the regression analysis 
         of $(Y,X,Z,U,V)$ on $(1,W, A ,B)$ for the glucose data}
\label{t:conc-5}
\end{table}

Additional numerical work not reported here has been carried out to
examine the sensitivity of the results to the choice of the point where
the MLE iteration sequence was stopped.
The overall conclusions are as follows: 
the regression coefficients and their observed significances are
stable over a wide range of stopping points; the individual components 
of  $\hat\alpha$ are not so stable, but the overall significance of the 
test for normality described at the end of Section~\ref{s:fit.msn}
remains well below 1\%.
The instability of the components of $\hat\alpha$ is not surprising 
considering that the sample size, $n=68$,  is small in this context, 
as discussed in Section~\ref{s:mle-weird}.

Reduction of the model, dropping components because of the 
non-significant coefficients or because of their irrelevance to the 
variables of interest, leads to consideration of the triplet $(Y,X,Z)$ 
with explanatory variables $(1,W,A)$. 
The new matrix $\hat\Omega^*$  and the vector $\hat\alpha$ are as reported 
in Table~\ref{t:conc-3}.

\begin{table}
\[
 \hat\Omega^*= \bordermatrix{
            &  Y    &  X      &    Z     \cr
      Y  & \M1.00  & -0.50  & \M0.00  \cr
      X  &  -0.50  &\M1.00  &  -0.52  \cr
      Z  & \M0.00  & -0.52  & \M1.00  }
\qquad
\hbox{\begin{tabular}{l r r r }
                 &  Y  &  X  &  Z  \\
\hline
$\hat\alpha$ &   2.50 & -21.42 &   -1.43\\
std. error   &   1.23 &   5.15 &    1.52 \\
$t$ ratio    &   2.04 &  -4.16 &   -0.94\\
\hline
\end{tabular}}
\]
\caption{\small\sl Matrix $\hat\Omega^*$, $\hat\alpha$ and other quantities 
         associated to the regression analysis 
         of $(Y,X,Z)$ on $(1,W, A)$ for the glucose data}
\label{t:conc-3}
\end{table}

The final graphical model has an edge between $(X,Y)$ and between 
$(X,Z)$ to represent conditional dependence, for fixed values 
of $(A, W)$ as indicated by $\hat\Omega^*$ and $\hat\alpha$ in
Table~\ref{t:conc-3}; 
background information can be used to choose a direction on these arcs.
Additional arcs are added from the fixed variables to the 
stochastic ones with the aid of the estimates and related $t$-ratios 
obtained from the last regression analysis,
namely directed arcs between   $(A,Y)$, and $(W,Z)$.

The pictorial representation of the graphical model is similar to
the regression graph of  Figure~6.4 of Cox \& Wermuth~(1996, p.\,141),
except for the arcs for they added on the basis of univariate regressions.
Clearly,  the building procedures and the associated interpretations
are a bit different,  and the two types of arcs (arising from conditional
dependence and from regression) should be kept graphically distinct
in our case.

We stress again that the above discussion intended to illustrate the
use of the conditional independence techniques with the aid of a well-known
dataset, not to produce a full data analysis. Moreover, the estimation 
method presented in Section~\ref{s:fit.msn} must be used with
caution with small samples like this one.

%===========================================================================

\section{An extension to elliptical densities} \label{s:elliptical}

The univariate skew-normal distribution was obtained by applying a 
skewing factor to the standard normal density, but the same method
is applicable to any symmetric density, as stated in Lemma~1 of
Azzalini~(1985).  This lemma can be extended to the $k$-dimensional
case where the notion of symmetric density is replaced by the notion
of elliptical density. 
% For a solid account on elliptical distributions, see Fang, Kotz \& Ng (1990).
The following lemma is a direct generalization
of Lemma~1 of Azzalini (1985), of which it also  follows the same line of 
argument in the proof.

\begin{lemma}  \label{th:skew}
Denote by $X$  a continuous random variable with density function $G'$ 
symmetric about 0 and by $Y=(Y_1,\ldots,Y_k)\T$ a continuous random 
variable with density function $f$, such that $X$ and $Y$ are independent. 
Suppose that the real-valued transform $W(Y)$
has symmetric density about 0.  Then
\begin{equation} \label{f:skew}
     \tilde{f}(y) = 2 \,f(y) G(W(y))
\end{equation}
is a $k$-dimensional density function.
\end{lemma}
\emph{Proof.} Since $X-W(Y)$ is symmetric about 0, then
\[ \half = \pr{X\leq W(Y)} = \E[Y]{\pr{X\leq W(Y)|Y}}
         = \int_{\Real^k} G(W(y))\,f(y)\,dy.  \]

\begin{corollary} 
Suppose that $X$ and $Y$ satisfy the conditions of the above lemma, and in 
addition that $Y$ has elliptical density centred at the origin; if   
\begin{equation} \label{f:W(y)}
  W(Y) = a_1 Y_1 + \cdots + a_k Y_k = a\T Y 
\end{equation}
then (\ref{f:skew}) is a $k$-dimensional density function for any choice
of $a$.
\end{corollary}
\emph{Proof.~}   The statement follows by noticing that $a\T Y$ 
has 1-dimensional elliptical distribution, 
i.e.\ its density is symmetric about 0.
See Theorem~2.16 of Fang, Kotz \& Ng (1990) for the distribution of a linear
transform of  elliptical variables.
\par\vskip 2ex\noindent

Clearly, (\ref{f:skew}) with $W(y)$ of type (\ref{f:W(y)}) 
includes the $\SN_k$ density for suitable choice of 
$f, G$ and any choice of $a$.

In principle, Lemma~\ref{th:skew} can be applied also to non-elliptical
densities.  For instance, if $Y\sim \SN_k$ and $a$ is
chosen suitably, according to  Proposition~\ref{th:lin1}, the density
of $W$ can be made normal, hence symmetric.  There is however a
major difference: in this case, the property holds for specific choices
of $a$ depending on the given choice of $f$, 
while with the elliptical densities it holds for all $a$'s.

Implicit in the proof of the lemma there is an acceptance--rejection idea,
hence a conditioning argument, similar to the one of Azzalini~(1986),
leading to the following method for random number generation.
If $X$ and $Y$ are as in above lemma, and
\[ Z  =\cases{Y    & if $X<W(Y)$, \cr
              -Y   & if $X>W(Y)$,  }
\]
then the density function of $Z$  is (\ref{f:skew}). 
In fact, its density  at point $z$ is
\[ f(z)G(W(z)) + f(-z) \{1-G(W(-z))\} \]
which is equal to $2\,f(z)\,G(W(z))$ if $f(z)=f(-z)$, a condition
fulfilled e.g.\ by elliptical densities centred at 0.

%  The lemma could be formulated in even more general forms. 
%  Consider $r$-dimensional rv's $X$ and $W(Y)$;
%  as long as we know 
%  \[ p=\pr{X_i\leq W(Y)_i \textrm{~for~} i=1,\ldots,r}, \] 
%  then we can use $1/p$ as the normalizing constant.
%  (Is  this extension of some practical relevance?)

%==============================================================================

\section{Further work}

Various issues related to the SN family have been discussed, but many
others remain pending.
Broadly speaking, these fall in two categories: open questions and 
further applications.

Among the open questions, the  anomalous behaviour of MLE in cases
described in section~\ref{s:mle-weird} is worth exploration even
\emph{per se}. 
In the multivariate case, construction of more accurate standard
errors would be welcome.
A more radical solution would be the introduction of the centred parametrization 
which has not been carried on from the univariate to the multivariate case.

Besides applications to  numerically more substantial
applied problems than those discussed here,  it is worth exploring the
relevance of the distribution in other areas of multivariate
statistics, in addition to those touched in section~\ref{s:multivar}.
A natural aspect to consider is the behaviour of other linear statistical 
methods outside normality, not only discriminant analysis.
Another relevant use could be in connection with sampling affected
by bias selection; this has been discusses by Copas \& Li (1997) and 
references quoted therein, in the case of a scalar response variable.
The skew-normal distribution offers the framework for a multivariate
treatment of the same problem, by consideration of its genesis via
conditioning.  

The  generalization to skew-elliptical densities has been left
completely unexplored. An adequate treatment of the connected
distributional and statistical issues requires the space of an
entire paper. Hence, this direction has not been explored here, but
a brief mention seemed to be appropriated, partly because of its
close connection with the SN distribution.

%==============================================================================

\vskip 2ex
\section*{Acknowledgments}
In the development of this paper, we much benefited from helpful and
stimulating  discussions with several colleagues.
Specifically, we are most grateful to John Aitchison for suggesting
the reparametrization adopted in subsection~\ref{s:fit.msn}, 
to Ann Mitchell for introducing us to elliptical densities, 
to Paul Ruud for discussions about the EM algorithm, 
to Monica Chiogna and David Cox for additional general discussions,
and to Samuel Kotz for his constant encouragement.
Additional extensive comments from the referees and the editor have
led to much better presentation of the material.
We also thank W.\,Q.\ Meeker for kindly providing the Otis data, and 
A.\,Albert for the liver data and associated informations.

A substantial part of this work has been developed while the first
author was at Nuffield College, Oxford, within the Jemolo Fellowship
scheme; the generous hospitality of the College is gratefully acknowledged.
Additional support has been provided by the `Ministero per l'Universit\`a
e per la Ricerca Scientifica e Tecnologica' and by `Consiglio Nazionale
delle Ricerche', Italy (grant no.\,97.01331.CT10).

%==============================================================================

\appendix
\section*{Appendices}

\subsection*{Two equivalent parametrizations}  \label{s:two-param}

We want to show that the $(\Omega, \alpha)$ parametrization adopted
in this paper is equivalent to the $(\lambda, \Psi)$ parametrization
of Azzalini \& Dalla Valle (1996).
 
The matrix $\Omega$ and the vector $\alpha$ appearing in (\ref{f:dens})  
were defined  in Azzalini \& Dalla Valle (1996) in terms
of a correlation matrix $\Psi$ and a vector 
$\lambda=(\lambda_1,\ldots,\lambda_k)\T$; specifically, they defined
\begin{eqnarray}
  \Delta&=& \diag\left(\recradice{1+\lambda_1^2},\ldots, 
                         \recradice{1+\lambda_k^2}\right), \label{f:Delta1}\\
  \Omega&=& \Delta(\Psi+\lambda \lambda\T)\Delta,          \label{f:Omega1} \\
  \alpha&=& \recradice{1+\lambda\T \Psi\inv\lambda}
                    \Delta\inv \Psi\inv \lambda   .        \label{f:alpha1}
\end{eqnarray}
Also, they defined $\delta=(\delta_1,\ldots,\delta_k)\T$
where $\delta_j=\lambda_j \recradice{1+\lambda_j^2}$ for $j=1,\ldots,k$.

With some algebraic work, it can  be shown that (\ref{f:Omega1})
and (\ref{f:alpha1}) are invertible, obtaining
\begin{eqnarray}  
  \Psi = \Delta\inv (\Omega -\delta\delta\T)\Delta\inv    \label{f:Psi} 
\end{eqnarray}
and (\ref{f:delta}), which then gives $\lambda$ using
$\lambda_j=\delta_j\recradice{1-\delta_j^2}$. As a by-product,
(\ref{f:alpha}) is obtained.

Clearly, for any choice of the $(\lambda,\Psi)$ pair, we obtain a
feasible $(\Omega,\alpha)$ pair; hence, we must only show the following.

\begin{proposition}
For any choice of the correlation matrix $\Omega$ and of the vector 
$\alpha\in\Real^k$, (\ref{f:dens}) is a density of $\SN_k$ type.
\end{proposition}
\emph{Proof.~} 
Given $\alpha$  and $\Omega$, compute $\delta$ using (\ref{f:delta}). 
This vector must satisfy condition $\Omega-\delta\delta\T\geq 0$, 
required by (\ref{f:Psi}); hence we must check that
\[ 
   \Omega - (1+\alpha\T \Omega\alpha)\inv \Omega\alpha\alpha\T\Omega
        \geq 0.
\] 
By using (\ref{f:smw}), the left-hand side can be seen to be equal to 
$(\Omega\inv+\alpha\alpha\T)\inv$ which is positive definite.
Moreover, fulfillment of $\Omega-\delta\delta\T\geq 0$ implies that 
all components of $\delta$ are less than 1 in absolute value.
Algebraic equivalence of (\ref{f:Delta1})--(\ref{f:alpha1})  and
(\ref{f:delta}),~(\ref{f:Psi}) completes the proof.

%==============================================================================

\subsection*{Gradient and hessian of the centred parameters}

The  partial derivatives of $\ell(CP)$ defined in Section~\ref{s:centred}
with respect to $(\beta,\sigma,\lambda)$ are
\begin{eqnarray*}
\pd{\ell}{\beta}  &=& (\sigma_z/\sigma)^2 X\T  \{ y-X\beta 
                      -\sigma\sigma_z\inv(\lambda p_1 - \mu_z 1_n)\}, \\
\pd{\ell}{\sigma} &=& -n/\sigma+\sigma_z(y-X\beta)\T(z-p_1\lambda)/\sigma^2,\\
\pd{\ell}{\lambda}&=& \frac{n}{\sigma_z} \sigma_z'
                      -z\T z' + p_1\T(z+\lambda z')
\end{eqnarray*}
where $z'$ denotes the derivative with respect to $\lambda$, and
\begin{eqnarray*}
  && p_1 =\zeta_1(\lambda z), 
    \qquad
     z' = \mu_z' + \sigma\inv(y-X\beta)\sigma_z' = \mu_z' + r \sigma_z',\\
  &&\mu_z' = \frac{\radice{2/\pi}}{(1+\lambda^2)^{3/2}},
    \qquad
    \sigma_z' = -\frac{\mu_z}{\sigma_z} \mu_z'.
\end{eqnarray*}
To obtain the partial derivatives with respect to $\gamma_1$, use
\[ 
  \pd{\ell}{\gamma_1} =\pd{\ell}{\lambda}/\der{\gamma_1}{\lambda}, 
  \qquad
  \der{\gamma_1}{\lambda} = \frac{3(4-\pi)}{2}\,
              \frac{\mu_z^2(\mu_z'\sigma_z-\mu_z\sigma_z')}{\sigma_z^4}.
\]
or equivalently 
\[ 
  \pd{\ell}{\gamma_1} =\pd{\ell}{\lambda}\:\der{\lambda}{\gamma_1}, 
  \qquad
  \der{\lambda}{\gamma_1} = 
      \frac{2}{3(4-\pi)} \left(\frac{1}{T\:R^2} + \frac{1-2/\pi}{T^3} \right)
\]
where
\[  
   R = \frac{\mu_z}{\sigma_z} = \left(\frac{2\gamma_1}{4-\pi}\right)^{1/3}
       \qquad
   T = \radice{2/\pi-(1-2/\pi)R^2}.
\]

The above derivatives lead immediately to the likelihood 
equations for $CP=(\beta,\sigma,\gamma_1)$. 
We need second derivatives for numerical efficient computations, and 
for computing the observed information matrix. The entries of the Hessian
matrix for $(\beta,\sigma,\lambda)$ are given by
\begin{eqnarray*}
 && -\pd{^2\ell}{\beta \partial\beta\T}  =  
      (\sigma_z/\sigma)^2 X\T(I_n-\lambda^2 P_2)X,  \\
 && -\pd{^2\ell}{\beta \partial\sigma}  =
      (\sigma_z/\sigma^2) X\T(z-\lambda p_1+(I_n-\lambda^2 P_2)(z-\mu_z1_n)),\\
 && -\pd{^2\ell}{\beta \partial\lambda} =
      \sigma\inv X\T\{\sigma_z'(-2r\sigma_z+\lambda p_1-\mu_z 1_n)
              +\sigma_z(p_1+\lambda P_2\tilde{z}-\mu_z'1_n)\}, \\
 && -\pd{^2\ell}{\sigma^2} =  
      \sigma^{-2}\{-n + 2\sigma_z r\T(z-\lambda p_1) +
       \sigma_z^2 r\T(I_n-\lambda^2 P_2)r\}, \\
 && -\pd{^2\ell}{\sigma \partial\lambda} = 
      -\sigma\inv r\T \left(\sigma_z'(z-\lambda p_1)+
            \sigma_z(z'-p_1-\lambda P_2\tilde{z}) \right),  \\
 && -\pd{^2\ell}{\lambda^2} =
       n\frac{(\sigma_z')^2-\sigma_z\sigma_z''}{\sigma_z^2} 
       + (z')\T z' + z\T z'' - \tilde{z}\T P_2\tilde{z}  
       -p_1\T (2z'+\lambda z'')
\end{eqnarray*}
where
\begin{eqnarray*}
  && r = \sigma\inv(y-X\beta), \qquad  \tilde{z} = z+\lambda z', \\
  && p_1=\zeta_1(\lambda z), \qquad P_2=\diag(p_2)=\diag(\zeta_2(\lambda z)), \\
  && z''= \der{z'}{\lambda} = \mu_z''+ \sigma_z''\sigma\inv(y-X\beta) , \\
  && \mu_z''= \der{\mu_z'}{\lambda} = -\frac{3\mu_z}{(1+\lambda^2)^2}, \qquad
     \sigma_z''= \der{\sigma_z'}{\lambda} = -\left(
                 \frac{\mu_z'(\mu_z'\sigma_z-\mu_z\sigma_z')}{\sigma_z^2} 
                  +\frac{\mu_z\mu_z''}{\sigma_z} \right) .
\end{eqnarray*}
Again, to obtain the Hessian matrix with respect  to $\gamma_1$ instead of 
$\lambda$, the last row and last column of the above  matrix
must be multiplied by $\diff{\lambda}/\diff{\gamma_1}$, except the
bottom right element which is computed as
\[
   -\pd{^2\ell}{\gamma_1^2} =  
     -\pd{^2\ell}{\lambda^2}  \left(\der{\lambda}{\gamma_1}\right)^2 
     - \pd{\ell}{\lambda} \left(\der{^2\lambda}{\gamma_1^2}\right).
\]
The final term of this expression is given by
\[  
  \der{^2\lambda}{\gamma_1^2} = -\frac{2}{3(4-\pi)}
     \left(
        \frac{T'}{(T\:R)^2}+\frac{2R'}{T\:R^3}+\frac{3(1-2/\pi)T'}{T^4}
     \right)
\]
where
\[
  R'= \der{R}{\gamma_1} = \frac{2}{3\:R^2(4-\pi)}, \qquad
  T'= \der{T}{\gamma_1} = -(1-2/\pi)\frac{R\:R'}{T}.
\]

For practical numerical work, the above quantities suffices.
If the expected Fisher information matrix $I_{CP}$ is needed, this is 
given by
% 
% Start from the information matrix for DP, namely
% \[
% I_{DP}= \pmatrix{
%          (1+\lambda^2 a_0)/\omega^2 & 
%          \{\mu_z(1+\mu_z^2 \pi/2)+\lambda^2 a_1\}/\omega^{2} &
%          \left(\dfrac{\radice{2/\pi}}{(1+\lambda^2)^{3/2}}-\lambda a_1\right)
%              /\omega
%        \cr
%           \{\mu_z(1+\mu_z^2 \pi/2)+\lambda^2 a_1\}/\omega^{2}&
%          (2+\lambda^2 a_2)/\omega^2 &
%         -\lambda a_2/\omega
%        \cr
%           \left(\dfrac{\radice{2/\pi}}{(1+\lambda^2)^{3/2}}-\lambda a_1\right)
%              /\omega &
%           -\lambda a_2/\omega &
%          a_2 }
% \]
% (essentially as given in SN85), where
% \[  a_k = a_k(\lambda) =\E{Z^k \zeta_1(\lambda Z)^2}, \qquad (k=0,1,2).  \]
% Then the information matrix for CP is given by
\[             I_{CP}  =  D\T\, I_{DP}\, D         \]
where $I_{DP}$ is the information matrix for the DP parameters, given
by Azzalini (1985) in the case $X=1_n$, and
\[  D = \left(\pd{(DP)_i}{(CP)_j}\right) 
      = \pmatrix{1   & -\dfrac{\mu_z}{\sigma_z}  & \pd{\xi}{\gamma_1}     \cr
                 0   &  \dfrac{1}{\sigma_z}      & \pd{\omega}{\gamma_1}  \cr
                 0   &   0                       & \pd{\lambda}{\gamma_1}  }
\]
where
\[
   \pd{\xi}{\gamma_1}     = -\frac{\sigma \mu_z}{3\sigma_z\gamma_1}, \qquad
   \pd{\omega}{\gamma_1}  = -\frac{\sigma\sigma_z'}{\sigma_z^2} 
                              \der{\lambda}{\gamma_1}.
\]

%==============================================================================

\section*{References}

\negindent
Aigner, D. J., Lovell, C. A. K. \& Schmidt, P. (1977).  Formulation
  and estimation of stochastic frontier production function model.
  {\em J. Econometrics} {\bf 12}, 21--37.
\negindent
Albert, A. \& Harris, E. K. (1987).
   \emph{Multivariate Interpretation of Clinical Laboratory Data}.
   Dekker, New York and Basel.
\negindent
Arnold, B.C., Beaver, R.J., Groeneveld, R.A. \& Meeker, W.Q. (1993).
   The nontruncated marginal of a truncated bivariate normal
   distribution. {\em Psychometrika} {\bf 58}, 471-478.
\negindent
Azzalini, A. (1985). A class of distribution which includes the normal ones.
   \emph{Scand. J. Statist.} {\bf 12}, 171--8.
\negindent
Azzalini, A. (1986). Further results on a class of distributions which 
   includes the normal ones. \emph{Statistica} {\bf 46}, 199--208.
\negindent
Azzalini, A. \& Dalla Valle, A. (1996). 
   The multivariate skew-normal distribution.
   \emph{Biometrika} {\bf 83}, 715--26.
\negindent
Barndorff-Nielsen, O. \& Bl{\ae}sild, P. (1983).
   Hyperbolic distributions. 
   In: \emph{Encyclopedia of Statistical Sciences}
   (ed.\ N.L.Johnson, S.Kotz \& C.B.Read), vol.\,3, 700--707.   
   Wiley, New York.
\negindent
Bl{\ae}sild, P. (1981).
   The two-dimensional hyperbolic distribution and related distributions,
   with an application to Johansen's bean data.
   \emph{Biometrika}, {\bf 68}, 251--63.
\negindent
Chiogna, M. (1997). Notes on estimation problems with scalar
   skew-normal distributions. Technical report 1997.15,
   Department of Statistical Sciences, University of Padua.
\negindent
Cartinhour, J. (1990). One dimensional marginal density function of a
  truncated multivariate Normal density function. {\em Comm.
  Statist., Theory and Methods} {\bf 19}, 197--203.
\negindent
Chou, Y.-M. \& Owen, D. B. (1984). An approximation to the percentiles
  of a variable of the bivariate normal distribution when the other
  variable is truncated, with applications. 
  \emph{Comm. Statist., Theory and Methods}, {\bf 13}, 2535--47.
\negindent
Cook, R.\,D. \& Weisberg, S. (1994). 
  {\em An Introduction to Regression Graphics}. Wiley,  New York.
\negindent
Copas, J. B. \& Li, H. G. (1997).
   Inference for non-random samples (with discussion).
   \emph{J.\,Roy.\, Statist.\, Soc.~B}, {\bf 59}, 55--95. 
\negindent
Cox, D, R, \& Wermuth, N. (1996).
   \emph{Multivariate dependencies: models, analysis and
    interpretation}. Chapman \& Hall, London.
\negindent
David, F. N., Kendall, M. G. \& Barton, D. E. (1966)
   \emph{Symmetric functions and allied tables}.
   Cambridge University Press.
   % segnatura 302-13 I
\negindent
Fang, K.-T., Kotz, S. \&  Ng, K. (1990).
   \emph{Symmetric multivariate and related distributions}.
   Chapman \& Hall, London.
\negindent 
Healy, M. J. R. (1968). Multivariate normal plotting.
   \emph{Appl. Statist.} {\bf17}, 157--161.
\negindent 
Johnson, N. L. \& Kotz, S. (1972).
   \emph{Distributions in statistics: continuous multivariate
   distributions}.  Wiley, New York
\negindent
Liseo, B. (1990).
   The skew-normal class of densities: Inferential aspects
   from a Bayesian viewpoint (in Italian).
   \emph{Statistica}, {\bf 50}, 59--70.
% \negindent
% Lancaster, P., Tismenetsky, M. (1985). \emph{The Theory of matrices,
% second edition, with applications}. Academic Press.
\negindent
Mardia, K. V. (1970).  Measures of multivariate skewness and kurtosis
   with applications. {\em Biometrika} {\bf 57}, 519--530.
\negindent
Mardia, K. V. (1974). Applications of some measures of multivariate skewness
   and kurtosis in testing normality and robustness studies.
   \emph{Sankhy\=a} {\bf B\,36}, 115-28.
\negindent
McCullagh, P. (1987). \emph{Tensor methods in statistics}.
   Chapman \& Hall, London.
\negindent
Meng, X.-L. \& van Dyk, D. (1997).
   The EM-algorithms --- an old folk-song sung to a fast new tune 
   (with discussion). \emph{J.Roy. Statist.\ Soc.\, B} {\bf 59}, 511--67.
\negindent
Muirhead, R. J. (1982).
   \emph{Aspects of multivariate statistical theory}.   Wiley, New York.
\negindent
Rao, C. R. (1947). 
  The problem of classification and distance between two populations.
  \emph{Nature}, {\bf 159}, 30--31.
\negindent
Rao, C.\,R. (1973). \emph{Linear statistical inference}, 2nd
   edition. Wiley, New York.
\negindent
Rotnitzky, A, Cox, D. R. Bottai, M. \& Robins, J. (1999).
   Likelihood-based inference with singular information matrix.
   To appear.
\end{document}

%% file: aa-macros.tex
\usepackage{amsfonts}
\newcommand{\anymode}[1]{\ifmmode{#1}\else\mbox{$#1$}\fi}

\newcommand{\negindent}{\par\vskip1ex\noindent\hangindent=2em \hangafter=1}
\newcommand{\NB}[1]{\texttt{/\textsuperscript{NB}}%
          \marginpar{\raggedright\sl\small#1\hfill}}	  

\long\def\Nota#1{\footnote{#1}\kern-0.2em\NB{cfr Nota$^{(\thefootnote)}$}}
\def\linea#1{\ifhmode\hfill\break\fi\hbox to \hsize{#1}}

\newcommand{\inv}{^{-1}}

\def\vc{v\kern -0.1em .c.\relax}

\newcommand{\dfrac}[2]{\displaystyle{\frac{#1}{#2}}}
\newcommand{\diag}{\mbox{\rm diag}}

\newcommand{\E}[2][]{
   \ensuremath{\mathbb{E}_{#1}\!\left\{\displaystyle{#2}\right\}}}

\newcommand{\half}{\mbox{$\textstyle \frac{1}{2}$}}   % p.172
\long\def\ignore#1{}

\newcommand{\indep}{\perp\kern-0.5em\perp}

\newcommand{\pr}[2][]{
   \ensuremath{\mathbb{P}_{#1}\!\left\{\displaystyle{#2}\right\}}}
\newcommand{\radice}[1]{\left(#1\right)^{1/2}}
\newcommand{\recradice}[1]{\left(#1\right)^{-1/2}}

\newcommand{\Real}{\mathbb{R}}

\newcommand{\tr}{\mbox{\rm tr}}
\newcommand{\T}{^{\top}}
\newcommand{\var}[2][]{
   \ensuremath{\textrm{var}_{#1}\!\left\{\displaystyle{#2}\right\}}}

\newtheorem{theorem}{Theorem}
\newtheorem{lemma}[theorem]{Lemma}

\newtheorem{corollary}[theorem]{Corollary}
\newtheorem{proposition}[theorem]{Proposition}
\newcommand{\SN}{\mathrm{SN}{}}

\newcommand{\diff}{\,\mathrm{d}}

\newcommand{\pd}[2]{\displaystyle\frac{\partial #1}{\partial #2}}